\documentclass[twocolumn]{aastex63}
\usepackage{natbib}
\usepackage{xcolor}
\usepackage{xspace,times,perpage}
\usepackage{graphicx}
\MakePerPage{footnote}

\newcommand{\lya}{Ly$\alpha$ }

\def \oii {[OII]}

\def \hi  H~{\sc i}
\newcommand{\hMpc}{\ensuremath{h^{-1}\,\mathrm{Mpc}}}

\newcommand{\msun}{${\rm M}_\odot$}

\begin{document}
\title{The Prime Focus Spectrograph Galaxy Evolution Survey}

\author{Jenny E. Greene}
\affiliation{Department of Astrophysical Sciences, Princeton University, Princeton, NJ 08544, USA}

\author{Rachel Bezanson}
\affiliation{Department of Physics and Astronomy, University of Pittsburgh, Pittsburgh, PA 15260, USA}

\author{Masami Ouchi}
\affiliation{Institute for Cosmic Ray Research, The University of Tokyo, 5-1-5 Kashiwanoha, Kashiwa, Chiba 277-8582,
Japan}
\affiliation{Kavli IPMU (WPI), The University of Tokyo, 5-1-5 Kashiwanoha, Kashiwa, Chiba 277-8583, Japan}

\author{John Silverman}
\affiliation{Kavli IPMU (WPI), The University of Tokyo, 5-1-5 Kashiwanoha, Kashiwa, Chiba 277-8583, Japan}
\affiliation{Department of Astronomy, School of Science, The University of Tokyo, 7-3-1 Hongo, Bunkyo, Tokyo 113-0033, Japan}

\author{St\'ephane Arnouts}
\affiliation{Aix Marseille University, CNRS, CNES, LAM, Marseille, France}
\author{Andy D. Goulding}
\affiliation{Department of Astrophysical Sciences, Princeton University, Princeton, NJ 08544, USA}
\author{Meng Gu}
\affiliation{Department of Astrophysical Sciences, Princeton University, Princeton, NJ 08544, USA}

\author{James E. Gunn}
\affiliation{Department of Astrophysical Sciences, Princeton University, Princeton, NJ 08544, USA}
\author{Yuichi Harikane}
\affiliation{Institute for Cosmic Ray Research, The University of Tokyo, 5-1-5 Kashiwanoha, Kashiwa, Chiba 277-8582, Japan}
\author{Timothy Heckman}
\affiliation{Johns Hopkins University}
\author{Benjamin Horowitz}
\affiliation{Department of Astrophysical Sciences, Princeton University, Princeton, NJ 08544, USA}
\author{Sean D. Johnson}
\affiliation{University of Michigan}
\author{Daichi Kashino}
\affiliation{Institute for Advanced Research, Nagoya University, Nagoya 464-8601, Japan}
\affiliation{Division of Particle and Astrophysical Science, Graduate School of Science, Nagoya University, Nagoya 464-8602, Japan}
\author{Khee-Gan Lee}
\affiliation{Kavli IPMU (WPI), The University of Tokyo, 5-1-5 Kashiwanoha, Kashiwa, Chiba 277-8583, Japan}
\author{Joel Leja}
\affiliation{Department of Astronomy \& Astrophysics, The Pennsylvania State University, University Park, PA 16802, USA} \affiliation{Institute for Computational \& Data Sciences, The Pennsylvania State University, University Park, PA, USA} \affiliation{Institute for Gravitation and the Cosmos, The Pennsylvania State University, University Park, PA 16802, USA}
\author{Yen-Ting Lin}
\affiliation{Institute of Astronomy and Astrophysics, Academia Sinica}
\author{Danilo Marchesini}
\affiliation{Tufts University}
\author{Yoshiki Matsuoka}
\affiliation{Research Center for Space and Cosmic Evolution, Ehime University}
\author[0000-0001-7457-8487]{Kentaro Nagamine} 
\affiliation{Theoretical Astrophysics, Department of Earth \& Space Science, Graduate School of Science, Osaka University,  1-1 Machikaneyama, Toyonaka, Osaka 560-0043, Japan} \affiliation{Kavli IPMU (WPI), The University of Tokyo, 5-1-5 Kashiwanoha, Kashiwa, Chiba 277-8583, Japan} \affiliation{Department of Physics \& Astronomy, University of Nevada, Las Vegas, 4505 S. Maryland Pkwy, Las Vegas, NV 89154-4002, USA}
\author{Yoshiaki Ono}
\affiliation{Institute for Cosmic Ray Research, The University of Tokyo, 5-1-5 Kashiwanoha, Kashiwa, Chiba 277-8582, Japan}
\author{Alan Pearl}
\affiliation{Department of Physics and Astronomy, University of Pittsburgh, Pittsburgh, PA 15260, USA}
\author{Takatoshi Shibuya}
\affiliation{Kitami Institute of Technology, 165 Koen-cho, Kitami, Hokkaido 090-8507, Japan}
\author{Michael A. Strauss}
\affiliation{Department of Astrophysical Sciences, Princeton University, Princeton, NJ 08544, USA}
\author{Allison L. Strom}
\affiliation{Department of Astrophysical Sciences, Princeton University, Princeton, NJ 08544, USA}
\author{Yuma Sugahara}
\affiliation{National Astronomical Observatory of Japan, 2-21-1 Osawa, Mitaka, Tokyo 181-8588, Japan} \affiliation{Waseda Research Institute for Science and Engineering, Faculty of Science and Engineering, Waseda University, 3-4-1, Okubo, Shinjuku, Tokyo 169-8555, Japan}
\author{Laurence Tresse}
\affiliation{Aix Marseille University, CNRS, CNES, LAM, Marseille, France}
\author{Kiyoto Yabe}
\affiliation{Kavli IPMU (WPI), The University of Tokyo, 5-1-5 Kashiwanoha, Kashiwa, Chiba 277-8583, Japan}
\author{Takuji Yamashita}
\affiliation{National Astronomical Observatory of Japan, 2-21-1 Osawa, Mitaka, Tokyo 181-8588, Japan}
\author{Masayuki Akiyama}
\affiliation{Astronomical Institute, Tohoku University}
\author{Metin Ata}
\affiliation{Kavli IPMU (WPI), The University of Tokyo, 5-1-5 Kashiwanoha, Kashiwa, Chiba 277-8583, Japan}
\author{Gabriel M. Azevedo}
\affiliation{Instituto de F\'isica, Universidade Federal do Rio Grande do Sul (UFRGS), Av. Bento Gon\c{c}alves, 9500, Porto Alegre, RS, Brazil}

\author{Chian-Chou Chen}
\affiliation{Academia Sinica Institute of Astronomy and Astrophysics (ASIAA), No. 1, Sec. 4, Roosevelt Road, Taipei 10617, Taiwan}
\author{Ana L. Chies-Santos}
\affiliation{Instituto de F\'isica, Universidade Federal do Rio Grande do Sul (UFRGS), Av. Bento Gon\c{c}alves, 9500, Porto Alegre, RS, Brazil}
\author{Richard S Ellis}
\affiliation{University College London}
\author{ChangHoon Hahn}
\affiliation{Department of Astrophysical Sciences, Princeton University, Princeton, NJ 08544, USA}
\author{Gabriel Roberto Hauschild Roier}
\affiliation{Universidade Federal do Rio Grande do Sul}
\author{Kohei Ichikawa}
\affiliation{Tohoku University}

\author{Kei Ito}
\affiliation{Department of Astronomy, School of Science, The University of Tokyo, 7-3-1 Hongo, Bunkyo-ku, Tokyo, 113-0033, Japan}
\author{Linhua Jiang}
\affiliation{Kavli Institute for Astronomy and Astrophysics, Peking University, Beijing 100871, China}
\author{Y.P. Jing}
\affiliation{Department of Astronomy, School of Physics and Astronomy, Shanghai Jiao Tong University, Shanghai, 200240, People’s Republic of China}
\author{Taiki Kawamuro}
\affiliation{Extreme Natural Phenomena RIKEN Hakubi Research Team, Cluster for Pioneering Research, RIKEN, 2-1 Hirosawa, Wako, Saitama 351-0198, Japan}
\author{Chiaki Kobayashi}
\affiliation{Centre for Astrophysics Research, Department of Physics, Astronomy and Mathematics, University of Hertfordshire, Hatfield, AL10 9AB, UK}
\author{Katarina Kraljic}
\affiliation{Aix Marseille University, CNRS, CNES, LAM, Marseille, France}
\author{Vincent Le Brun}
\affiliation{Aix Marseille University, CNRS, CNES, LAM, Marseille, France}
\author{Xin Liu}
\affiliation{Department of Astronomy, University of Illinois at Urbana-Champaign, 1002 W. Green Street, Urbana, IL 61801, USA; National Center for Supercomputing Applications,  1205 W. Clark Street, Urbana, IL 61801, USA}
\author{Peter Melchior}
\affiliation{Department of Astrophysical Sciences, Princeton University, Princeton, NJ 08544, USA}
\author{Claudia L. Mendes de Oliveira}
\affiliation{Departamento de Astronomia, Instituto de Astronomia, Geof\'isica e Ci\^encias Atmosf\'ericas da USP, Cidade \\ Universit\'aria, 05508-900, S\~ao Paulo, SP, Brazil}
\author{Rog\'erio Monteiro-Oliveira}
\affiliation{Academia Sinica, Institute of Astronomy and Astrophysics, 11F of AS/NTU Astronomy-Mathematics Building, No.1, Sec. 4, Roosevelt Rd, Taipei 10617, Taiwan, R.O.C.}
\author{Tohru Nagao}
\affiliation{Ehime University}
\author{Masato Onodera}
\affiliation{Subaru Telescope, National Astronomical Observatory of Japan, National Institutes of Natural Sciences (NINS), 650 North A'ohoku Place, Hilo, HI 96720, USA}
\author{Roderik~A. Overzier}
\affiliation{Observat\'{o}rio Nacional/MCTI, Rua General Jos\'{e} Cristino, 77, S\~{a}o Crist\'{o}v\~{a}o, Rio de Janeiro, RJ 20921-400, Brazil}
\author{David Schiminovich}
\affiliation{Department of Astronomy, Colunbia University}
\author{Malte Schramm}
\affiliation{Planetary Exploration Research Center, Chiba Institute of Technology, Tsudanuma, Narashino, Chiba 275-0016, Japan}
\author{Rhythm Shimakawa}
\affiliation{National Astronomical Observatory of Japan}
\author{Laerte Sodr\'e}
\affiliation{Departamento dAstronomia, IAG, Universidade de S\~ao Paulo}
\author{Thaisa  Storchi-Bergmann}
\affiliation{Instituto de Fisica - Universidade Federal do Rio Grande do Sul, RS, Brazil}
\author{Tomoko L. Suzuki}
\affiliation{Kavli IPMU (WPI), The University of Tokyo, 5-1-5 Kashiwanoha, Kashiwa, Chiba 277-8583, Japan}
\author{Masayuki Tanaka}
\affiliation{National Astronomical Observatory of Japan, 2-21-1 Osawa, Mitaka, Tokyo 181-8588, Japan}
\author{Yoshiki Toba}
\affiliation{National Astronomical Observatory of Japan}
\author{Hideki Umehata}
\affiliation{Nagoya University}
\author{Jiaqi Wang}
\affiliation{Department of Astronomy, School of Physics and Astronomy, and Shanghai Key Laboratory for Particle Physics and Cosmology, Shanghai Jiao Tong University, Shanghai 200240, China}
\author{Charlotte Welker}
\affiliation{Johns Hopkins University}
\author{Katherine  Whitaker}
\affiliation{University of Massachusetts Amherst}
\author{Kun Xu}
\affiliation{Department of Astronomy, School of Physics and Astronomy, Shanghai Jiao Tong University, Shanghai, 200240, People’s Republic of China}
\author{Yongquan Xue}
\affiliation{CAS Key Laboratory for Research in Galaxies and Cosmology, Department of Astronomy, University of Science and Technology of China, Hefei 230026, China}
\author{Xiaohu Yang}
\affiliation{Department of Astronomy, School of Physics and Astronomy,  Shanghai Jiao Tong University, Shanghai 200240, China}
\author{Ying Zu}
\affiliation{Department of Astronomy, School of Physics and Astronomy, Shanghai Jiao Tong University, Shanghai 200240, China}

\date{\today}


\begin{abstract}
    We present the Prime Focus Spectrograph (PFS) Galaxy Evolution pillar of the 360-night PFS Subaru Strategic Program. This 130-night program will capitalize on the wide wavelength coverage and massive multiplexing capabilities of PFS to study the evolution of typical galaxies from cosmic dawn to the present. From Lyman $\alpha$  emitters at $z \sim 7$ to probe reionization, drop-outs at $z\sim3$ to map the inter-galactic medium in absorption, and a continuum-selected sample at $z \sim 1.5$, we will chart the physics of galaxy evolution within the evolving cosmic web. This article is dedicated to the memory of Olivier Le Fevre, who was an early advocate for the construction of PFS, and a key early member of the Galaxy Evolution Working Group.
\end{abstract}

\hspace{5mm}

\section{Introduction}

\begin{figure*}[!t]
    \centering
    \includegraphics[width=\textwidth]{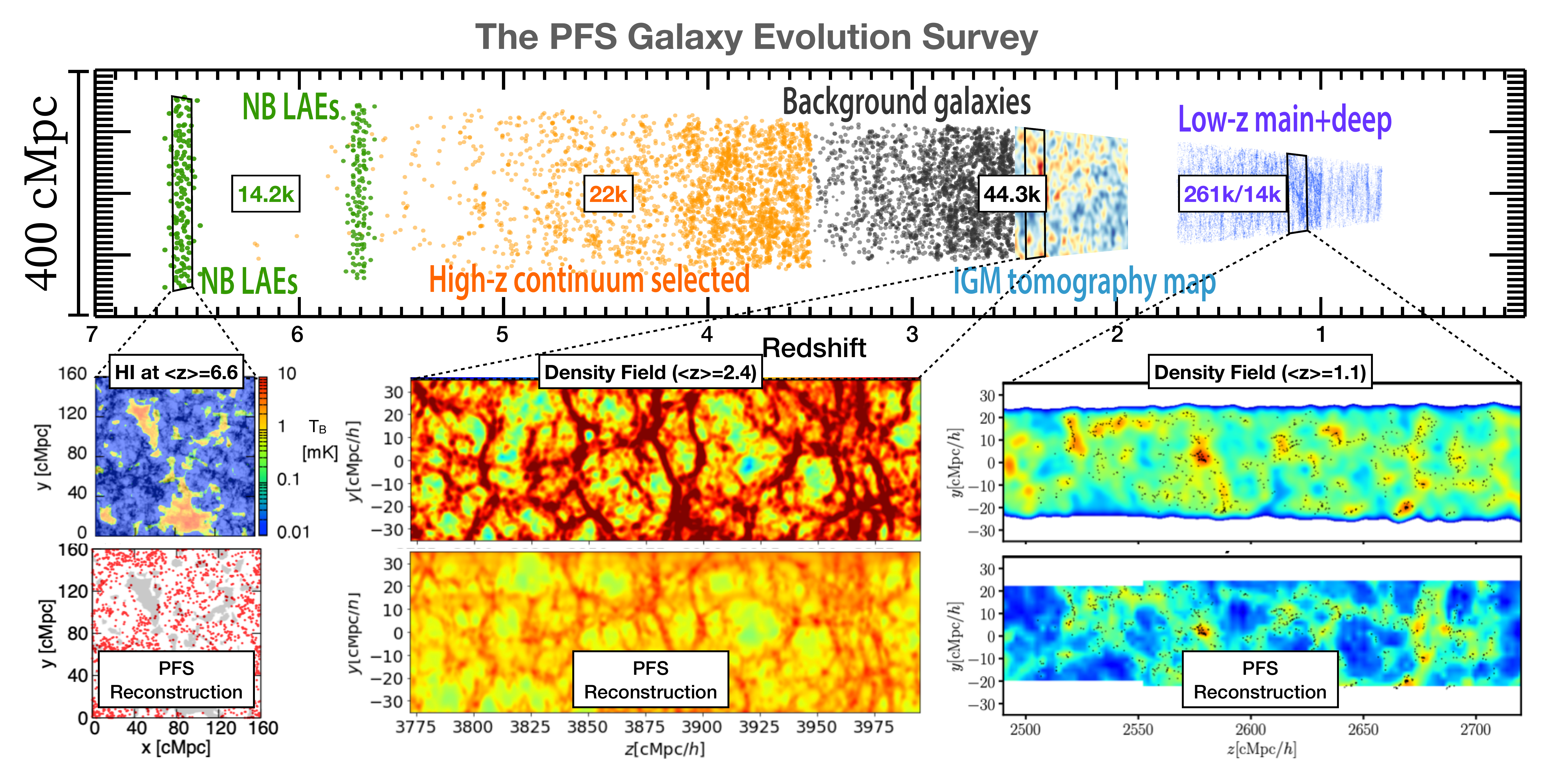}
    \caption{Number of spectra in each subset of the PFS GE survey (top) and a representative light cone to demonstrate the key redshift regimes it will probe (middle). Pop-out panels depict the recovery of cosmic structures from the PFS SSP (top panels: truth, bottom panels: recovered) at 
$\langle z\rangle=1.1, 2.4$ and 6.6.
The top left panel shows reionization bubbles in the HI 21cm brightness temperature distribution from 
\citet{Kubota2019},
while the bottom panel shows the LAE distribution observed by PFS spectroscopy, which is anti-correlated with the ionized bubbles in this particular model. The top middle and right panels show the simulated density fields from the Horizon AGN simulation at 
$\langle z\rangle = 2.4$ and $\langle z\rangle =1.1$, 
while bottom middle and right show the reconstructed density from the PFS galaxy redshift distribution and IGM absorption data using the ARGO \citep{Ata::2015} and TARDIS \citep{Horowitz2019} algorithms, respectively.
}
    \label{fig:LSS}
\end{figure*}

The evolution of galaxies is inextricably linked to the cosmic web \citep{Somerville15}.
After inflation, the primordial density fluctuations grow via gravitational instability, increasing the density contrast in the dark matter, and eventually forming a cosmic web of sheets, filaments, and nodes containing virialized dark matter halos. Baryons flow with the dark matter on large scales, and some are incorporated into the halos. Unlike dark matter, baryons can lose energy by radiation, and sink deeper into the potential well. In the simplest picture, this inflow is halted by centrifugal forces and the baryons form a disk  \citep{Mo:1998,Burkert2016}.
Stars form within the central regions of these disks, as do the seeds of supermassive black holes.  Galaxies continue to grow over billions of years, primarily through continuing accretion of gas from the web, and secondarily through mergers with other dark matter haloes and their baryonic contents \citep[e.g.,][]{Vogelsberger:2014}.

In this standard Cold Dark Matter paradigm, the formation and evolution of galaxies are driven both by the collapse, growth, and assembly of dark matter halos in which galaxies reside, and by complex baryonic physics. In particular, feedback from massive stars and supermassive black holes play a critical role 
\citep[e.g.,][]{croton:2006}. The large-scale environment, on a variety of scales, may also impact both the dark matter accretion history and baryonic processes in complex ways \citep[e.g.,][]{Cooper:2008,Peng2010,Kovac::2014}. To fully map the connection between large-scale structure and galaxy evolution, we need to explore a large range of scales, from kpc to tens of Mpc, and redshifts from $z=1$ to 6, which can be mapped with a dense spectroscopic survey covering more than 10 deg$^2$ in order to cover transverse scales of $\gtrsim 100$~Mpc. 

This is the primary objective of the Galaxy Evolution Survey planned for the Prime Focus Spectrograph, as we describe below. The paper will proceed as follows. In \S \ref{sec:cosmicweb} we lay out the main science questions that drive the survey design, while in \S \ref{sec:surveydesign}, we describe the samples required to answer these questions. \S \ref{sec:photometry} presents the available photometry to select these samples and \S \ref{sec:mock} describes the sensitivity assumptions and mock galaxy catalogs that we use for planning. Finally, in \S \ref{sec:deliverable} we justify the survey design by presenting the deliverables that we have extracted from our mock spectra. Throughout this document we use the AB magnitude system unless specified otherwise.

\begin{figure*}[t]
    \centering
    \includegraphics[width=\textwidth]{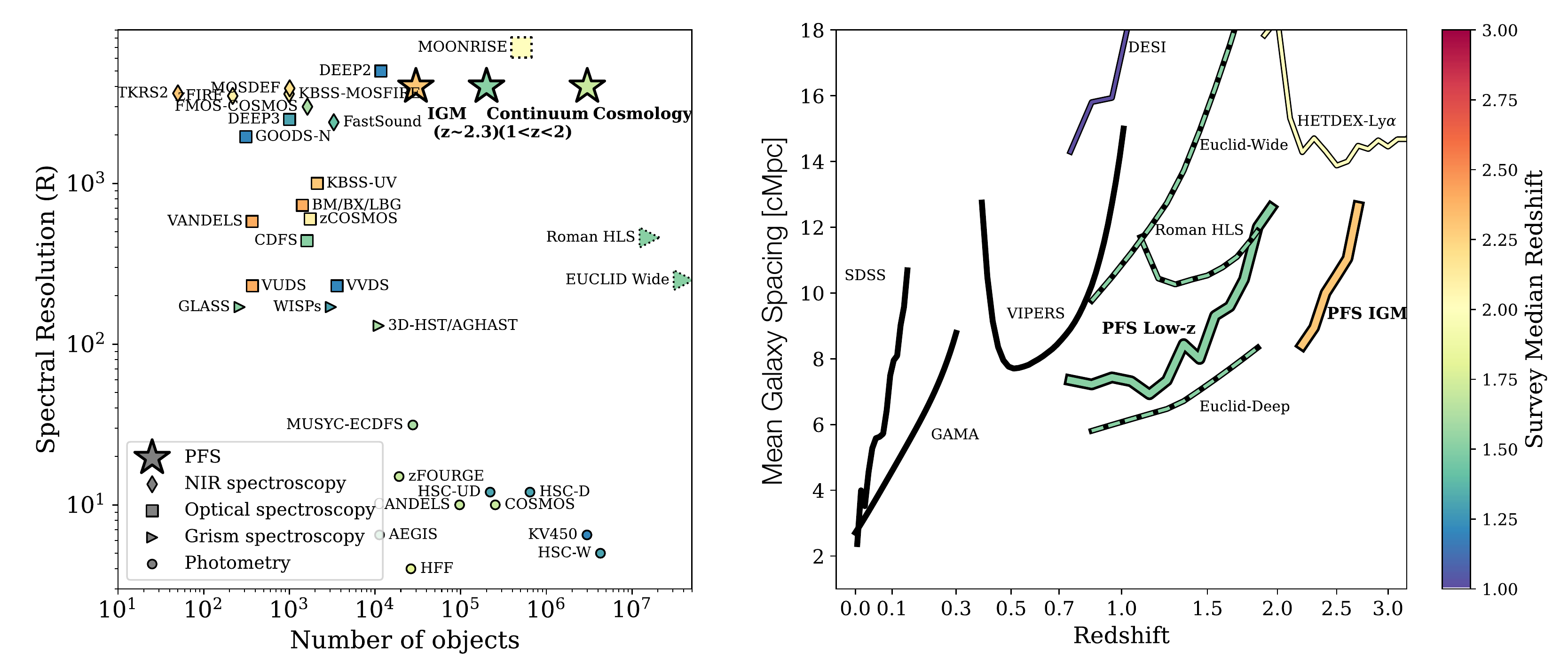}
    \caption{(Left) The combined spectral resolution and statistical samples produced by the PFS GE survey components would lie in unprecedented parameter space at $z \gtrsim 1$ (compilation adapted from \citealt{forsterschreiber:21}). (Right) The dense sampling and large volume probed by the PFS sample will extend maps of the cosmic web beyond the local Universe. Our high sampling reaches smaller average inter-galaxy separations than previous high redshift surveys.}
    \label{fig:pfsvothers}
\end{figure*}

\section{Half a Million Spectra Covering 6 Gyrs of Cosmic Evolution}
\label{sec:intro}

Outstanding questions about the physics of galaxy formation cannot be answered without a large spectroscopic survey at $z > 1$. Spectra provide a rich suite of diagnostics of the fundamental properties of galaxies and their environments. The sample must be large to map out the distributions of, and causal connections between, these properties, and to chart their evolution in redshift. In order to optimally measure the galaxy properties, the spectral resolution must be matched to the internal velocity dispersions of galaxies, and the signal-to-noise ratio must be sufficient to measure the key spectral diagnostics. To connect the properties of galaxies to the large-scale cosmic web, the spectroscopic survey must have sufficient spatial sampling over a wide enough area to average over cosmic variance and span environments at density extremes.

\setlength{\tabcolsep}{1pt}
\begin{deluxetable}{lcccc}[!h]
\tabletypesize{\footnotesize}
\tablecaption{Galaxy samples and depths}
\tablehead{\colhead{Type}&
\colhead{Redshift}&
\colhead{Selection}&
\colhead{Exp. Time}&
\colhead{\# of spectra} \vspace{-0.3cm} \\
\colhead{} &\colhead{range} & \colhead{} & \colhead{(hrs)} & \colhead{(${\times}10^3$)}}
\startdata
Continuum &$0.7 - 2$&$y, J{<}22.8$&2, 12& $261,\, 14$ \\
IGM&$2.1 - 3.5$&$y{<}24.3$,$g {<} 24.7$&6, 12 & $30.3, \, 14$ \\
LBG&$3.5 - 7$&$y{<}24.5$ & 6 & 22 \\
LAE& $2.2, \, 5.7, \, 6.6$ & \, L$_{\rm Ly\alpha}{>}3{\times}10^{42}$ erg s$^{-1}$  & $3, \, 6, \, 12$ & $7.4, \, 4.5, \, 2.8$ \\
AGN-GE&$0.5-6.0$&various (see text)&$1-5$&4.2
\enddata
\tablecomments{Shortened target table, including the main science targets, their redshift ranges, the main selection criteria (in AB mags), exposure times, and expected number of spectra. ``Continuum''  refers to the $0.7{<}z{<}1.7$ survey; ``IGM'' is the inter-galactic medium tomography  More details are given in \S \ref{tbl:target} below.}
\label{tbl:basicsamples}
\end{deluxetable}

The Subaru Prime Focus Spectrograph (PFS), currently under construction for the 8.2m Subaru Telescope on Maunakea, Hawai'i, is specifically designed to make these goals feasible. With 2394 fibers covering a 1.25 deg$^2$ field-of-view and wide wavelength coverage spanning the ultraviolet to the near-infrared ($0.38-1.25\mu$m), the PFS is uniquely positioned in the coming decade to probe the evolution of typical Milky Way-like galaxies from the epoch of reionization at $z \sim 7$ to the present. The PFS Galaxy Evolution (GE) survey will target complementary sub-samples, each designed to fully leverage the instrumental capabilities and capture the physical properties of galaxies at critical moments in cosmic history (see Table \ref{tbl:basicsamples} and Figure \ref{fig:LSS}). The GE survey will leverage deep imaging from the Hyper Suprime-Cam Subaru Strategic Program (HSC-SSP; \citealt{Aihara2018}) Deep fields, as described in detail in \S \ref{sec:photometry}. The PFS Galaxy Evolution survey will be one pillar of a proposed 360-night Subaru Strategic Program for PFS (PFS-SSP) following commissioning. Together, the three pillars -- Cosmology, Galactic Archeology, and Galaxy Evolution -- will focus on {\it Cosmic Evolution and The Dark Sector}; the Galaxy Evolution pillar is a critical component that focuses on tying evolution in dark matter halos with galaxies \citep{takada:2014}.

The PFS Galaxy Evolution survey represents a dramatic improvement in sample size, spectral resolution, and density of targeting over previous spectroscopic surveys of the critical epoch of galaxy formation ($z{>}1$). Previous studies have been limited to lower redshifts [e.g., DEEP-2 \citep{newman:2013}, zCOSMOS \citep{lilly:2007}, VVDS \citep{lefevre:2005}, LEGA-C \citep{vanderwel:2016}] or at higher-z to much smaller, biased samples  [e.g., KBSS \citep{steidel:2014}, MOSDEF \citep{kriek:2015}, and FMOS-COSMOS \citep{Silverman:2015}] or low spectral resolution (e.g., 3D-HST, \citealt{Momcheva:2016}) (Fig. \ref{fig:pfsvothers}). Our survey is also highly complementary to the other upcoming massively multiplexed spectroscopic survey studying the properties of $z =1-6$ galaxies, MOONRISE \citep{Maiolino2020}. VLT/MOONS will cover the rest-frame optical lines to $z \approx 2$, allowing for better characterization of inter-stellar medium (ISM) physics at cosmic noon. PFS will have more than twice the number of fiber-hours ($\sim$1.3 million versus 500,000 for the MOONRISE XSwitch strategy) and blue coverage. These together allow us to perform an inter-galactic medium (IGM) tomography survey, characterizing the cosmic web and the IGM at cosmic noon, go deep on continuum-selected galaxies at $z \approx 1.5$, and probe Lyman $\alpha$ (Ly$\alpha$) emission in galaxies at $z\gtrsim2$. Finally, our high spectral resolution and broad wavelength coverage will also be complementary to the Euclid \citep{Laureijs:2011} and Roman Space Telescope \citep{Spergel:2015} grism surveys. 

We will study the peak epoch of star formation using continuum-selected galaxies with $J{<}22.8$, which will capture $\sim$90\% of the $M_*~{\gtrsim}~3 {\times} 10^{10}$~\msun~population. This Main sample of 300,000 galaxies will have a ${\sim}70\%$ average completeness and thus include multiple galaxies in groups down to $M_{\rm halo} {\sim} 10^{13.5}$~\msun. Exposure times of 2-hr integrations will provide a high spectroscopic redshift completeness. We will integrate for 8-12 hrs on an additional 14,000 galaxies (the ``Deep'' sample) to measure stellar ages, chemical abundance ratios, stellar velocity dispersions, and faint emission lines such as [O~III]$\, \lambda \, 4363$. 

The overarching connection of galaxies to large-scale structures will be extended out to $z{\sim}6$. In particular, the distribution of neutral hydrogen in the cosmic web at $z{=}2.1{-}2.5$ 
will be mapped at a co-moving scale of $\sim$4~Mpc through an IGM tomographic experiment based on the detection of Ly$\alpha$ absorption seen in the spectra of background galaxies at $2.5{<}z{<}3.5$. From $3.5{<}z{<}7$, we will target Lyman Break Galaxies (LBGs) to $y{<}24.5$, which will provide a powerful sample for studying the early formation of galaxies and their clustering strength. These will be complemented at the furthest distances with a sample of $\sim$15k \lya\ Emitters (LAE) at $6{<}z{<}7$ selected from the Subaru/HSC narrow-band imaging to probe the formation of young galaxies at cosmic dawn.

Together, these samples will allow us to jointly address two main themes from $z \sim 1$ to the epoch of reionization. On the largest scale, we will map the cosmic web through the distribution of galaxies (\S\ref{sec:cosmicweb_evolving} and \S \ref{sec:cosmicweb_stat}). On smaller scales, we will measure the evolution of the properties of the stars and gas in galaxies to elucidate the underlying physical processes driving their formation and growth (\S \ref{sec:internal}). 

\setlength{\tabcolsep}{15pt}
\begin{deluxetable}{lc}[!h]
\tabletypesize{\footnotesize}
\tablecaption{Large Scale Structure}
\tablehead{\colhead{Component of the Web}&
\colhead{Expected Number}}
\startdata
$M_{\rm halo}>10^{13} \, M_{\odot}$ & 2200 \\
$M_{\rm halo}>10^{13.5} \, M_{\odot}$ & 450 \\
$M_{\rm halo}>10^{14} \, M_{\odot}$ & 35 \\
Voids ($z{<}2$, $r > 7$ cMpc) & 132,000 \\
Voids ($z{<}2$, $r > 20$ cMpc) & 3,000 \\
Voids ($z>2$, $r > 7$ cMpc) & 1000 \\
Protoclusters ($2{<}z{<}6$) & 100 
\enddata
\tablecomments{Top three rows quantify the number of massive halos that we expect above this limit in the PFS volume. Next three lines quantify the number of voids with the given redshift and size limits. The final entry is the expected number of protoclusters.}
\label{tbl:lss}
\end{deluxetable}

\section{Galaxies within the Cosmic Web}
\label{sec:cosmicweb}

The growth of structure drives the evolution of dark matter halos and the flow of gas between and into galaxies, and therefore is the fundamental process behind the formation and evolution of galaxies themselves.  On the largest scales, we have strong theoretical reasons to believe that galaxies evolving in voids will have different star formation histories and angular momentum distributions from those in filaments or nodes. We know that the orientation of filaments does impact the spins of galaxies \citep[e.g.,][]{Keres2005,Pichon2011,Zhang2013}, and possibly their star formation histories \citep[e.g.,][]{Kraljic2018}, while mergers which take place along filaments can change the galaxy angular momentum and may lead to quenching \citep[e.g.,][]{Dubois2014}. Large galaxy redshift surveys like VIPERS and zCOSMOS at $z \approx 0.7$ are just starting to detect these effects \citep[][]{Malavasi2017,Laigle2018}. Prior to PFS it was not possible for a single survey to test this at earlier times with simultaneously (a) wide enough area to probe the rarest overdensities and overcome cosmic variance and (b) dense enough sampling to trace the large-scale structure on ${\sim}1{-}3$~Mpc scales. Thanks to our unprecedented multiplexing and wavelength coverage, we will perform the redshift survey needed to make a high-fidelity map of the large-scale structure in the distant Universe and situate the galaxy properties within that context. 

\subsection{Charting the Evolving Cosmic Web}
\label{sec:cosmicweb_evolving}

PFS will substantially expand on existing surveys both in sky volume and object density, mapping galaxies and gas at three critical epochs. We show the key targets for cosmic web reconstruction in Figure \ref{fig:LSS}. With the Main and Deep samples at $z \sim 1.1$ we will test the role of the cosmic web in regulating star formation at cosmic noon (right). At $z\sim 2.4$, we will use IGM tomography to produce the most detailed and largest-scale maps of the intergalactic medium that fuels galaxies at the peak epoch of star formation (center). Finally, the program will use LAEs at $z = 6-7$ to map the reionization of the Universe (left). Together, these three regions will provide a representative picture of the galaxy-environment relationship over cosmic time. 

We will spectroscopically identify unprecedented numbers of groups, (proto)clusters, and voids of a range of sizes (see Table \ref{tbl:lss}) to facilitate unprecedented tests of environmental effects on galaxy evolution at early times (see more details in \S \ref{sec:group_photo}). We will explore the role of galaxy overdensity in driving star formation histories and quenching as a function of mass and redshift \citep[e.g.][]{Peng2010,Wang2018,Lemaux2022}. Deviations between galaxy environment and the associated HI absorption will highlight the growth of the warm-hot intergalactic medium \citep{cen05a} and intracluster medium \citep{kooistra:2021}. 

\subsection{Statistically connecting Galaxies to the Cosmic Web}
\label{sec:cosmicweb_stat}

\begin{figure}[t]
    \centering
    \includegraphics[width=0.44\textwidth]{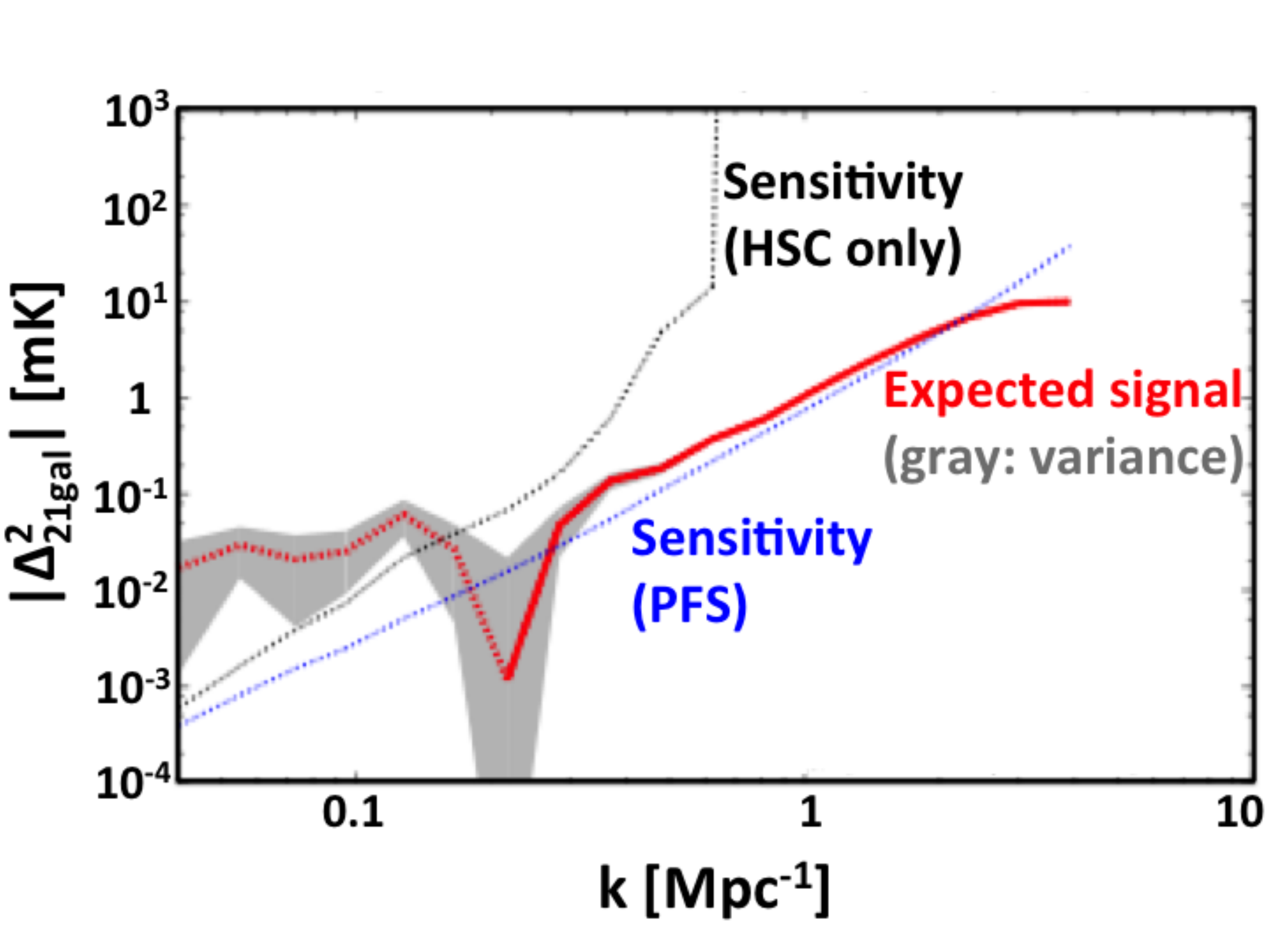}
    \caption{Spatial distribution of LAEs and imprint on reionization: The absolute cross-power spectrum of brightness temperature between LAEs and 21cm emission at $z=6.6$ ($|\Delta^2_{\rm 21gal}|$) is shown as a function of the wavenumber $k$. The red solid line indicates the expected signal predicted by numerical simulations tailored for the PFS survey \citep{Kubota:2018}. The red dotted line is the same as the red solid line, but indicates the negative signal, and the cross-power spectrum turnover is found at $k\sim 0.2$ Mpc$^{-1}$. The grey shade represents the variance in the PFS survey field. The black and blue dotted lines denote the sensitivity of LAE observations accomplished by the HSC SSP survey alone and the combination of HSC+PFS observations, respectively.}
    \label{fig:LAEx21}
\end{figure}

At $5.5 {<} z {<} 7$, spanning the epoch of reionization, we will map the galaxy-cosmic web connection by measuring the spatial cross-power spectrum between the
ionizing galaxies, detected by PFS as LAEs,
and the \ion{H}{1} gas in the cosmic web detected in redshifted 21 cm
emission by the SKA1 array (Key Science Project observations start in 2024; Fig. \ref{fig:LAEx21}). 
Models predict that on small scales (single
ionized bubbles), one-halo clustering introduces a positive correlation. Beyond a bubble radius, an anti-correlation will result if reionization proceeds from regions of high to low density. The amplitude
of the signal, the spatial scale at the correlations become negative (the cross-power spectrum transition), and the overall shape of the cross-power spectrum, all constrain
the reionization history of the Universe \citep[e.g.,][]{Sobacchi:2016}.

With the PFS LAE spectroscopic data, we will calculate the cross-power spectrum of the spatial distribution of the LAEs and redshifted {\sc Hi} 21-cm radio emission using the data from the SKA-1 at $z\sim 6-7$ (Hasegawa et al. 2016), aiming at the first detection of the epoch of reionization 21-cm emission and the identification of the cross-power spectrum transition at the $3-5\sigma$ levels. The radiation transfer models of \citet{Kubota:2018} demonstrate that the PFS LAEs and the 21-cm data will allow us to detect the cross-power spectrum turnover (Fig.\,\ref{fig:LAEx21}), which cannot be accomplished with only HSC imaging, due to the redshift uncertainty of LAEs. The PFS measurements for the spatial distribution of LAEs will reveal the cross-power spectrum turnover scale ($k\sim 0.2$ Mpc$^{-1}$) with the accuracy of $\delta k=0.1$ and the overall shape of the cross-correlation, which determine the reionization history and ionized bubble topology, respectively.

At cosmic noon, the IGM tomography sample will not only trace the cosmic web, but also give us new insight into the state of the gas in and around galaxies. A critical component of our program is the measurement of foreground galaxy redshifts (i.e., 25k continuum-selected galaxies, 9.2k LAEs, and 1.3k AGN) that lie within the \ion{H}{1} web. With this sample, we will use galaxy morphological features identified through HSC (and later Roman) imaging to constrain possible intrinsic alignments with respect to the cosmic web traced by the HI tomography \citep{krolewski17a}. We will also measure the 3D cross-correlation between the galaxies and HI absorption to constrain the underlying bias (and hence halo mass) of the galaxies as a function of stellar mass, star-formation rate, metallicity, and other properties.  Cross-correlations can also be extended to metals reflecting the velocity field near galaxies, allowing the study to extend down to CGM scales in conjunction with detailed hydrodynamical simulations \citep{Turner17,Nagamine21}.

With PFS, we can robustly measure the evolution of the stellar mass function and the distribution of specific star-formation rates as a function of the local mean over-density, to extend known trends between quenching and environment at $z{<}1$. There are tantalizing clues that the sign of the morphology--density relation may change at higher redshift, with strong star formation occurring in protocluster cores at $z\gtrsim 2$ \citep[e.g.][]{wang2016}.
We will also test whether the 3D location in the web (e.g. distance from the nearest filament) plays an additional role in affecting galaxy properties over a range of epochs
\citep{Laigle2018}.

\begin{figure}
    \centering
    \includegraphics[width=0.45\textwidth]{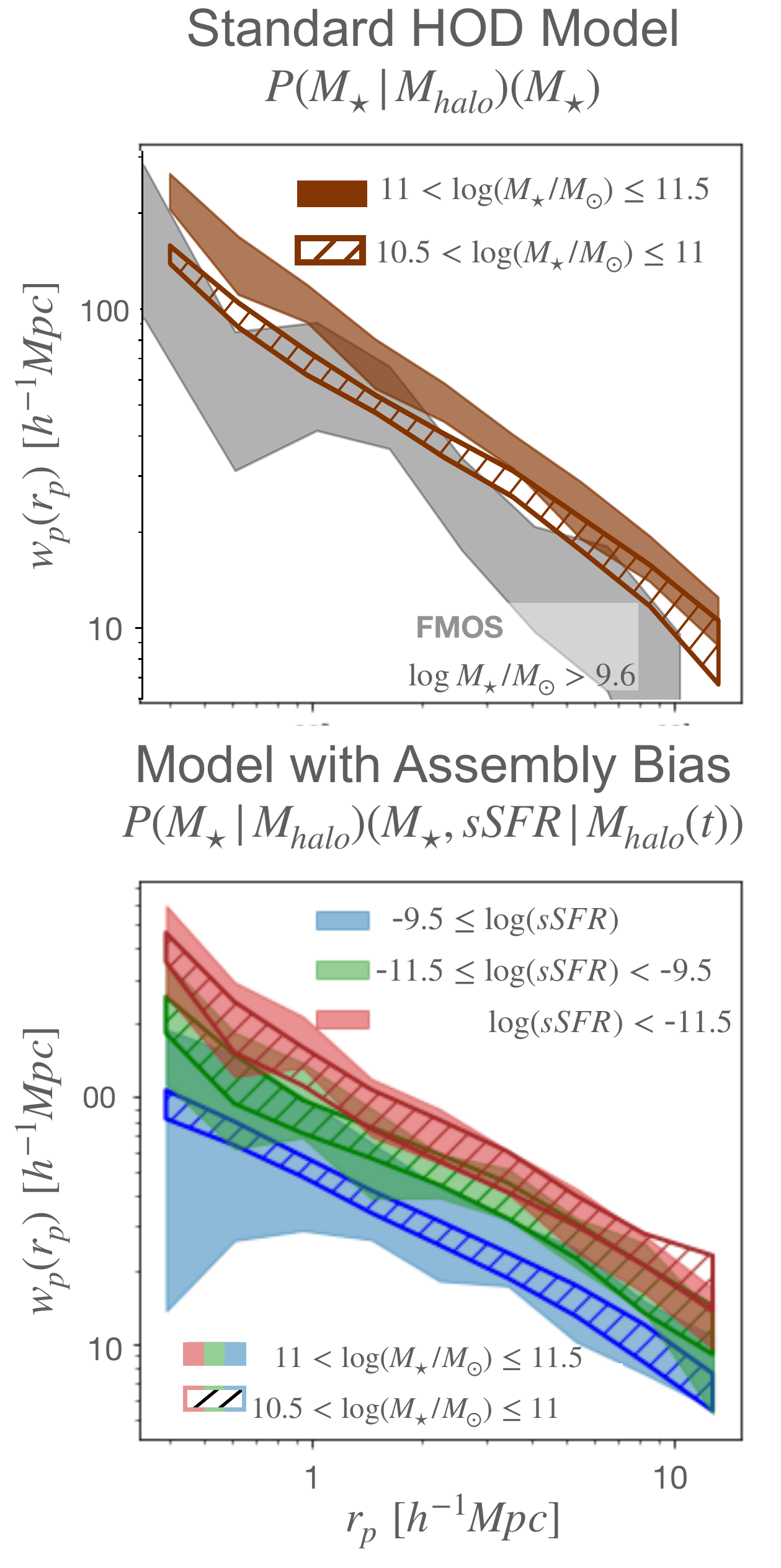}
    \caption{The expected projected clustering signal of continuum-selected galaxies as measured from the PFS mock \citep{pearl:22}, with errors characteristic of the planned continuum-selected ($z{<}2$) sample volumes after subdividing on stellar mass and specific star formation rate, without (top) and with (bottom) assembly bias to show that the PFS Survey will allow unprecedented clustering measurements, versus e.g., the FMOS survey (gray) that can only measure clustering above a mass threshold \citep{kashino:2017}. Two-point correlation functions of spectroscopic samples of galaxies provide a robust statistical measurement of host dark matter halo mass, thus constraining the SMHM relation; PFS will constrain this relation from $0.7\lesssim z\lesssim 4.5$. Furthermore, we will test models of the overall efficiency of galaxy formation, as well as the existence of assembly bias in the mapping from galaxy stellar mass to halo mass. }
    \label{fig:efficiency}
\end{figure}

On $\sim$Mpc scales, the connection between galaxies and their host dark matter halos has provided a compelling framework to understand the overall efficiency of galaxy formation \citep[e.g.][]{weschler:2018}. The most basic measure of this relation is the stellar mass-to-halo mass (SMHM) relation \citep[e.g.,][]{Yang2012}, which captures the overall efficiency of star formation over the entire history of the Universe \citep[e.g.][]{Behroozi13, Yang2013}. This relation is often derived using two-point statistics to compare the biased clustering of galaxies at a given stellar mass to compute the average masses of host dark matter haloes (Figure \ref{fig:efficiency}a). The default analytic models used to describe the ``galaxy-halo connection" rely on deterministic mappings (including scatter) between $M_{halo}$ and $M_{\star}$ \citep[e.g.,][]{berlind:02, Yang2003}, but increasing evidence suggests that additional secondary factors, such as relative halo assembly history, are important in driving the timing and efficiency of galaxy formation. Empirically, this ``assembly bias'' manifests as stronger clustering of older (or less-star-forming) galaxies at fixed stellar mass (Fig. \ref{fig:efficiency}b) \citep[e.g.,][]{Gao:07}. No other photometric or spectroscopic survey at $z>1$ would have the necessary statistics, accurate 3D positions, stellar masses, star formation rates, and halo masses derived from clustering measurements and halo occupation distribution models \citep[e.g.,][]{Durkalec2015,Kashino2017} to measure the SMHM relation and test for the importance of assembly bias in understanding the galaxy-halo connection in the early Universe.  

Finding clusters, voids, and filaments from $0.7{{<}}z{{<}}3$ demands the following survey design parameters: (a) area $>$10 deg$^2$ to probe $>1$ Gpc$^3$ cosmic volumes and cover the largest overdensities, (b) sampling ${>}70\%$ to get multiple galaxies in groups to 10$^{13.5}$~\msun, and (c) IGM background targets to i$=$24.7 mag to sample the IGM with 3~Mpc resolution.

\subsection{The Evolution of Internal Galaxy Properties: \\From Statistics to Physics}
\label{sec:internal}

To fully understand the drivers of galaxy evolution, we need to simultaneously determine both what is flowing into galaxies (accretion of gas, dark matter, and galaxies) and the galactic winds driven by star formation and black holes. We need an accurate accounting of the star formation and chemical evolution histories of galaxies at each epoch.

Our survey will have unprecedented wavelength coverage for half a million galaxies from $ 0.7\lesssim z \lesssim 6$, which will allow us to quantify the strength and prevalence of neutral outflows (\S \ref{sec:gas_flows}), identify and study the properties of active galactic nuclei that may also drive outflows, and situate all of these physical properties within the evolving cosmic web. We can then match the gas flows with the star formation histories of individual galaxies (\S \ref{sec:sf_quenching}) and chemical abundances (\S \ref{sec:chemical_evolution}).

\subsubsection{Gas Flows}
\label{sec:gas_flows}

Most galaxies grow primarily through the accretion of gas from the cosmic web. This inflow fuels new star formation and black hole growth, which in turn can drive outflows that may change the dynamical and thermal state of the inflowing gas. This cycle likely plays a key role in the self-regulation (and quenching) of star formation and black hole growth. PFS will chart the evolution of intergalactic gas on large scales, starting from reionization (as probed by LAEs) through the metallicity and spatial distribution of the IGM at $z {\sim }2$ with the IGM tomography experiment. 

With PFS, feedback from both massive stars and AGN can be characterized by measuring the incidence rate and outflow velocities of warm ionized gas as traced by blue-shifted inter-
stellar absorption-lines. This will require stacking to achieve the necessary S/N, but the PFS sample size is so large that the stacking can be done in many bins in the 4-D space of redshift, stellar mass, star formation rate (SFR), and AGN luminosity. This will enable direct tests of competing feedback models: momentum-driven \citep[e.g.][]{murray:05} vs.\ energy-driven winds \citep[e.g.][]{chevalier:85}, that predict how outflow velocities should depend on these quantities.

\begin{figure*}
    \centering
    \includegraphics[width=0.85\textwidth]{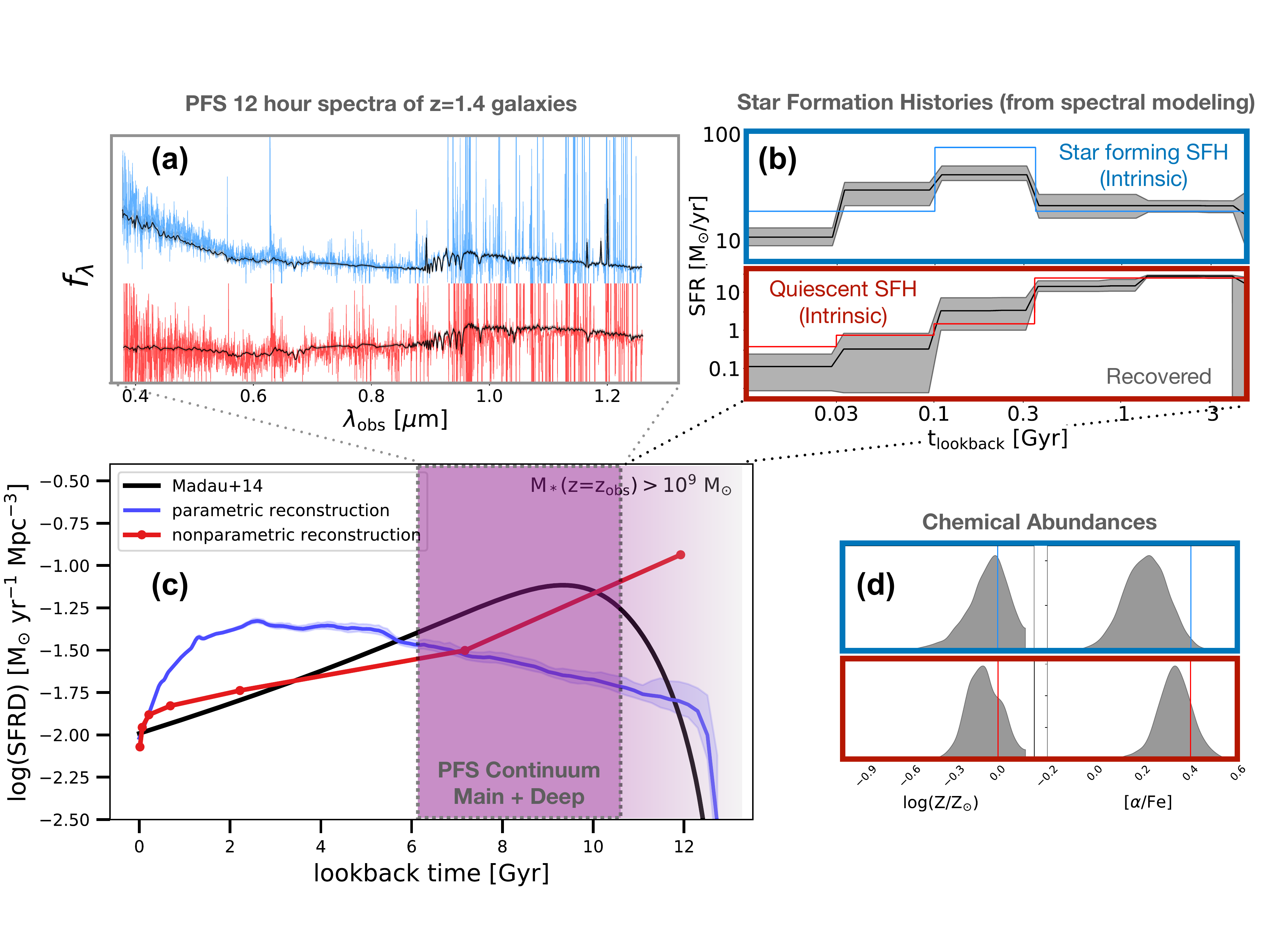}
    \caption{Demonstration that PFS spectra will recover star formation histories over the key epoch of cosmic noon. In (d) we show that star formation histories inferred from local galaxy spectra (red and blue lines) fail to capture the basic rise-and-fall of the history of star formation (black), and particularly diverge at cosmic noon \citep[adapted from][]{leja:19}. On the other hand, PFS spectroscopy demonstrated in (a) from the Deep sample will directly probe (${\lesssim}2$Gyr) star formation histories at this crucial epoch, capturing both higher order moments (e.g. quenching, bursts) for individual galaxies and the average star formation history of the galaxy population to high precision. Specifically, in (b) we show models (red and blue lines) representing non-parametric reconstructions that manage to recover the main features of the input star formation histories, including a galaxy with a recent burst hidden under a fresh phase of star formation (top) and a galaxy which quenches slowly and gradually over several Gyr (bottom). Finally, in (d) we see that [Fe/H] and, crucially, nonsolar abundances captured via [Mg/Fe] are recovered within the uncertainties as well.}
    \label{fig:SFHs}
\end{figure*}

The interface where the inflow meets feedback is the circum-galactic medium (CGM), a region with a scale of roughly the virial radius, and which contains a baryonic mass comparable to the stellar component \citep{tumlinson:2017}. PFS will probe the CGM around the $z {\sim} 1$ continuum-selected sample using the Ly$\alpha$ tomography sample as backlights, reaching expected equivalent widths of 0.3~\AA\ within 200 kpc impact parameters for 100-galaxy stacks. The galaxy sample from the PFS cosmology survey will also be probed by SDSS quasars as backlights, reaching 0.10~\AA\ equivalent widths in stacks in bins of redshift and impact parameter. At the same time, as described in the next two sections, we will have access to the detailed star formation and chemical enrichment histories of the target galaxies to connect with their CGM reservoirs and larger scale environments.

\subsubsection{Charting Detailed Star Formation Histories}
\label{sec:sf_quenching}

The bulk of star formation over cosmic time occurs in galaxies on the ``Star Forming Main Sequence'' \citep[e.g.,][]{noeske:2007}, a correlation between SFR and  $M_\ast$ that evolves strongly towards higher SFR with increasing redshift 
\citep[e.g.,][]{Speagle:2014}. While this evolution can be largely understood as tracing the overall cosmic rate of accretion and merging, the processes that lead to a relatively small dispersion in the relation at a given redshift are poorly understood.  Additional populations lie above and below this relation; starburst galaxies are responsible for $\sim$10\% of the global star formation and a population of quenched galaxies dominate at the highest masses today. 

With the PFS dataset, the distribution of galaxies along and across the main sequence, and the fraction of star-forming, starburst, quenching, and quenched galaxies will be determined as a function of redshift. For the subset with deep (12~hour) exposures, the continuum spectra will hold critical information to describe higher-order moments of the star formation histories of galaxies, where photometric measurements are fundamentally limited by the adopted priors \citep[e.g.][]{carnall:19}, leading to the power of PFS to reconstruct real star-formation histories at cosmic noon (Figure \ref{fig:SFHs}). In particular, we can expect to reconstruct reliable star-formation histories and recover reliable abundance ratios, for individual galaxies at $0.7{<}z{<}1.7$, particularly from the Deep spectra (see also \S \ref{sec:deliversfr}). Thus, we will begin to quantify the extent of ongoing star formation, the importance of rejuvenation, and time the quenching for $>$10,000 galaxies between $0.7{<}z{<}2$. 

There is good evidence that the quenching process has evolved significantly between $z\sim0$ and 1 \citep{Behroozi18}.
As described in Sec.~\ref{sec:cosmicwebmap}, PFS will provide a detailed mapping between environmental metrics on scales from individual dark matter halos to large scales (e.g., filaments and voids) and galaxy star formation histories. We will also explore the role of active galactic nuclei (AGN) in halting star formation. The main sample of AGN will be drawn directly from the galaxy sample using a variety of strong-line criteria \citep[e.g.,][]{baldwin:1981,juneau:2011}, making it straightforward to quantify incidence rates in the 3D parameter space of stellar mass, SFR, and redshift \citep{Kauffmann:2004,Silverman:2009}.
This sample will be supplemented by AGN selected by a diverse set of multi-wavelength criteria, designed to fully probe the population of AGN at these redshifts.

\subsubsection{Chemical Evolution}
\label{sec:chemical_evolution}

The chemical abundances of galaxies provide powerful constraints on their prior star-formation histories and on the roles of inflows of low-metallicity gas and outflows of high-metallicity gas \citep{tremonti:2004,Maiolino2019}. Historically, the state-of-the-art for studying chemical abundances of galaxy populations has been the measurement of the correlation between galaxy stellar mass and metallicity \citep{garnett2002,erb2008,finlator2008}. With PFS, the statistical characterization of the stellar mass--gas-phase metallicity relation at $z{\sim}0.7-1.7$ will for the first time be comparable to the analysis that has been possible for almost two decades at $z\sim0$ using SDSS, while at the same time PFS Deep spectra could provide the first stellar mass--stellar metallicity relation beyond the local universe \citep{Gallazzi:2005,Kirby:2013}. This represents a significant advance over what is currently possible with smaller and more heterogeneously-selected samples at intermediate redshifts \citep[$z{<}1$; e.g.,][]{lamareille2009,zahid2011}. Because of the dramatic changes in galaxies' star formation histories during this period, completing a detailed, robust census of enrichment as a function of galaxy properties will serve as an important benchmark for theoretical predictions.

In addition to single element abundances like O/H or Fe/H, PFS will also enable the determination of stellar and gas-phase abundance ratios such as O/Fe, N/O, and C/O with sufficient S/N. Iron abundances can be determined from stellar continua in deep individual spectra or in spectral stacks, or by inferring the shape of the dominant ionizing radiation field from an ensemble of emission lines. Both N and C are primarily measured using emission lines in individual galaxy spectra: the [\ion{N}{2}]$\lambda\lambda6549,83$ doublet can only be observed with PFS at $z\lesssim0.9$, but beyond $z>1$ the \ion{C}{3}]$\lambda\lambda1907,9$ doublet will probe  gas density in higher-ionization gas. The ratio of stellar $\alpha$ abundances to Fe that we can recover from the Deep spectra (Fig.\ \ref{fig:SFHs}) probe the most recent episode of star formation, distinguishing between contributions of Type Ia supernovae from low-mass stars with that of Type II from high-mass stars. C and N are mainly formed in low- and intermediate-mass stars, respectively, tracing different timescales from core-collapse and Type Ia supernovae \citep[note that the timescale of C production is comparable to SNIa;][]{Vincenzo:2019}. With PFS, this analysis can be performed with stacked deep spectra in bins of stellar mass, SFR and, uniquely, with redshift.

We foremost rely on the wide wavelength coverage and high spectral resolution of PFS to model detailed galaxy physics from our spectra. We require deep integrations (12 hrs) to unlock detailed star formation histories for individual or small ensembles of galaxies, as well as to detect the weak emission lines ([O~III]4363, C~III]) that are required to model the chemical enrichment history of the gas. Likewise, deep integrations are required to reach high target density (and thus faint targets) to map the metallicity of the CGM.

\section{Galaxy Samples}
\label{sec:surveydesign}

\begin{figure*}
\begin{center}
\includegraphics[angle=0,width=0.85\textwidth]{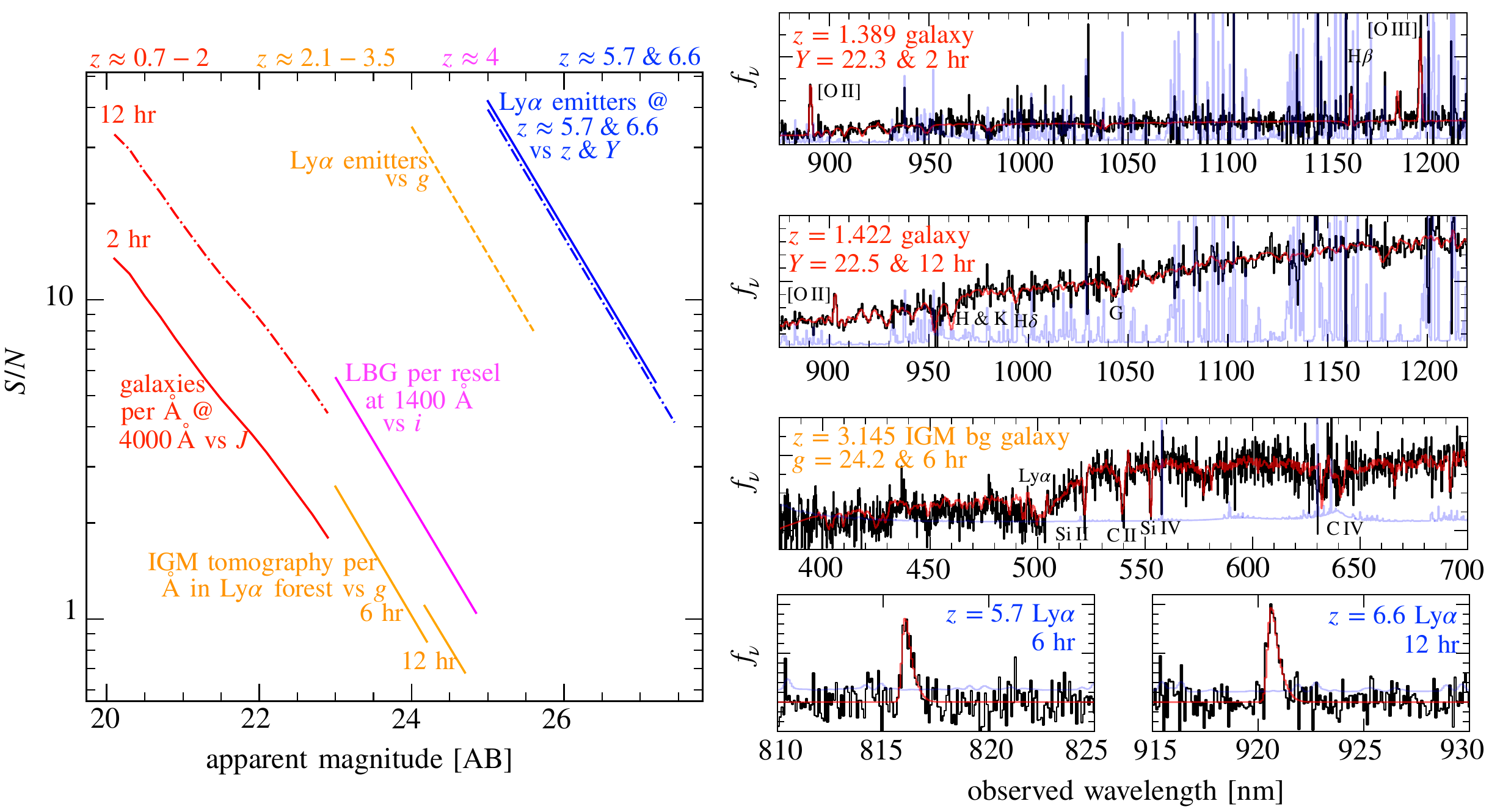}
\figcaption[]{{\it Left}: Spectral S/N achieved as a function of apparent magnitude for different components of the survey, based on the simulator described in \S \ref{sec:mock}. Different samples are evaluated at different observed wavelengths, as indicated. LBGs at $z \approx 4$ are shown per resolution element. {\it Right}: Example mock PFS spectra with the integration times as indicated. We show a typical galaxy spectrum observed for two (top) and 12 (second) hours (\S \ref{sec:lowz}). We also show the Ly$\alpha$ region of a $z=3.145$ star-forming galaxy (third), from which we will measure the Ly$\alpha$ absorption from gas at $z \sim 2.1-2.5$ (\S \ref{sec:igm}). At the bottom we show two example LAEs (\S \ref{sec:lae}).
} 
\label{fig:snrspectra}
\end{center}
\end{figure*}

We summarize the main samples here, along with key science drivers motivating the area and sampling choices. We note that the number of required fiber hours (FH), the product of the exposure time per object by the number of spectra, is meaningful unit in which to compare time investments rather than nights, since we will be interleaving programs with different target selections throughout for efficiency. 

\subsection{Lyman $\alpha$ emitters}
\label{sec:lae}

We will target a total of 12k LAEs at $z\sim 2$ and $6-7$ with $L\gtrsim L_*$ selected from the Subaru/HSC narrowband (NB) imaging data \citep{ouchi2018,shibuya2018a,shibuya2018b,konno2018}. \citet{shibuya2018b} have obtained spectroscopic confirmation of $z=5.7$ and $6.6$ LAEs with $\simeq 1-1.5$ hr integrations using Subaru/FOCAS and Magellan/LDSS3, while subsequent near-IR (NIR) spectroscopy with Subaru/nuMOIRCS and Keck/MOSFIRE show detections of associated UV nebular lines.

We use numerical simulations of LAEs whose spectra with Ly$\alpha$ emission are determined by radiative transfer calculations that are tuned to reproduce observed LAE samples in a partially neutral IGM \citep{inoue2020}. Based on simulations of the PFS spectra  (\S \ref{sec:deliverism}), we set the on-source PFS integration time to 6 (12) hours, allowing us to detect Ly$\alpha$ at $z\sim 6-7$ down to 
$\log L_{\rm Ly\alpha}/[{\rm erg\ s^{-1}}]>42.7$ $(42.5-42.7)$ at the $\gtrsim 5 \sigma$ level. 

Furthermore, we consider the detectability of strong UV nebular lines, observed ratios of nebular UV lines \citep{Stark2015a,Stark2015b,Sobral2015,Stark2017} for sources with $\log L_{\rm Ly\alpha}/[{\rm erg\ s^{-1}}]>43.0$. We set the on-source PFS integration time to 3 hours for $z\sim 2$ LAEs with $\log L_{\rm Ly\alpha}/[{\rm erg\ s^{-1}}]>42.5$. Assuming the average Ly$\alpha$ to {\sc [Oii]} line ratio of \citet{erb2016}, we expect to detect {\sc [Oii]} lines in those $z \sim 2$ LAEs with a luminosity $\log L_{\rm Ly\alpha}/[{\rm erg\ s^{-1}}]>43.0$. In total, we request 68k FH  (Table \ref{tbl:target}).

\subsection{Drop-out galaxies ($2 {<} z {<} 7$)}
\label{sec:igm}

The primary drivers of the survey design at $z>2$ are (1) the tomographic mapping of diffuse HI gas as traced by \hbox{Ly$\alpha$} absorption, (2) the spatial correlation between HI gas and galaxies, and (3) the two-point projected correlation function in multiple bins of redshift and spatial scale. These will be met by targeting 54k galaxies spread over the 12.3 square degrees of the PFS GE program. Targets will be selected from the HSC Deep layer using five broad-band filters ($grizy$) and CLAUDS $u$-band imaging \citep{Sawicki2019}. The $y$-band limit (24.3 or 24.5; see Table~\ref{tbl:target}) ensures a high degree of success ($\sim70\%$) with measuring spectroscopic redshifts while an additional $g$-band selection is required to probe the foreground IGM for galaxies at $2.5{<}z{<}3.5$ (see below). Exposure times are nominally 6 or 12 hours, in order to provide the S/N that is required by the IGM tomography program. In total, this sample requires 395k FH of observing time.

This spectroscopic sample will provide accurate redshifts to build upon the remarkable LBG samples available from the HSC Deep and Ultradeep layers. \citet{Harikane2022} have used 4.1 million LBGs at $2 {<} z {<} 7$ to construct their rest-UV luminosity functions and angular correlation functions which demonstrate the rise of the AGN population at high luminosities ($M_{UV}\lesssim-23$). Thus, the PFS-SSP will enable studies of the connection between massive galaxies, supermassive black holes, and environment in the early universe for the first time with sufficient statistical accuracy.

\subsubsection{Area, Sampling, Depth for IGM Tomography}

The primary requirement for the {\sc Hi} Ly$\alpha$ tomographic mapping is the reconstruction of the 3D gas distribution using background separations of $\langle d_{\perp}\rangle \approx 4$\,cMpc.
\citet{lee14b} have demonstrated that 8--10m optical telescopes can effectively provide spectra (S/N $\geq$ 2 per resolution element of $\sim 2.5$\AA) in a reasonable amount of time of the required density of sight lines.
Based on simulations \citep[see Figure 8 in][]{lee14b} and previous
observations \citep{lee14a}, we estimate that observing 18.3k galaxies with $2.5{<}z{<}3.5$ and $g{<}$24.7 would yield the required density of sight lines ($\sim$1600 deg$^{-2}$), that are separated by $\langle d_{\perp}\rangle \approx 2.5 \hMpc$ , to produce reconstructed maps at an effective map S/N$\approx$3 per $2.5 \hMpc$ side-length volume element.   In practice, we divide the sample into  bright and faint components, splitting at $g=24.2$, with different exposure times.  If we were to limit ourselves to the bright objects, the sight line separation would increase to $3.7 \hMpc$. 

The exposure times for the bright and faint components are set by the $g$-band limiting magnitudes of 24.2 (t$_{\rm exp}$=6\,hrs) and 24.7 (t$_{\rm exp}$=12\,hrs). These exposure times are required to reach at least S/N=2 per 1.8\mbox{\AA} resolution element, evaluated at
around 4200\mbox{\AA} in the observed frame. Additional continuum-selected targets provide a redshift-space distribution of galaxies in the same cosmic volume as the HI map, and allow us to investigate the HI--galaxy correlation. In total, the HI tomography component of the survey requires 287k FH including the foreground galaxies at $2.1{<}z{<}2.5$.

\begin{figure}
\centering
\vbox{ 
\vskip -5mm
\hskip -1mm
\includegraphics[angle=0,width=0.35\textwidth]{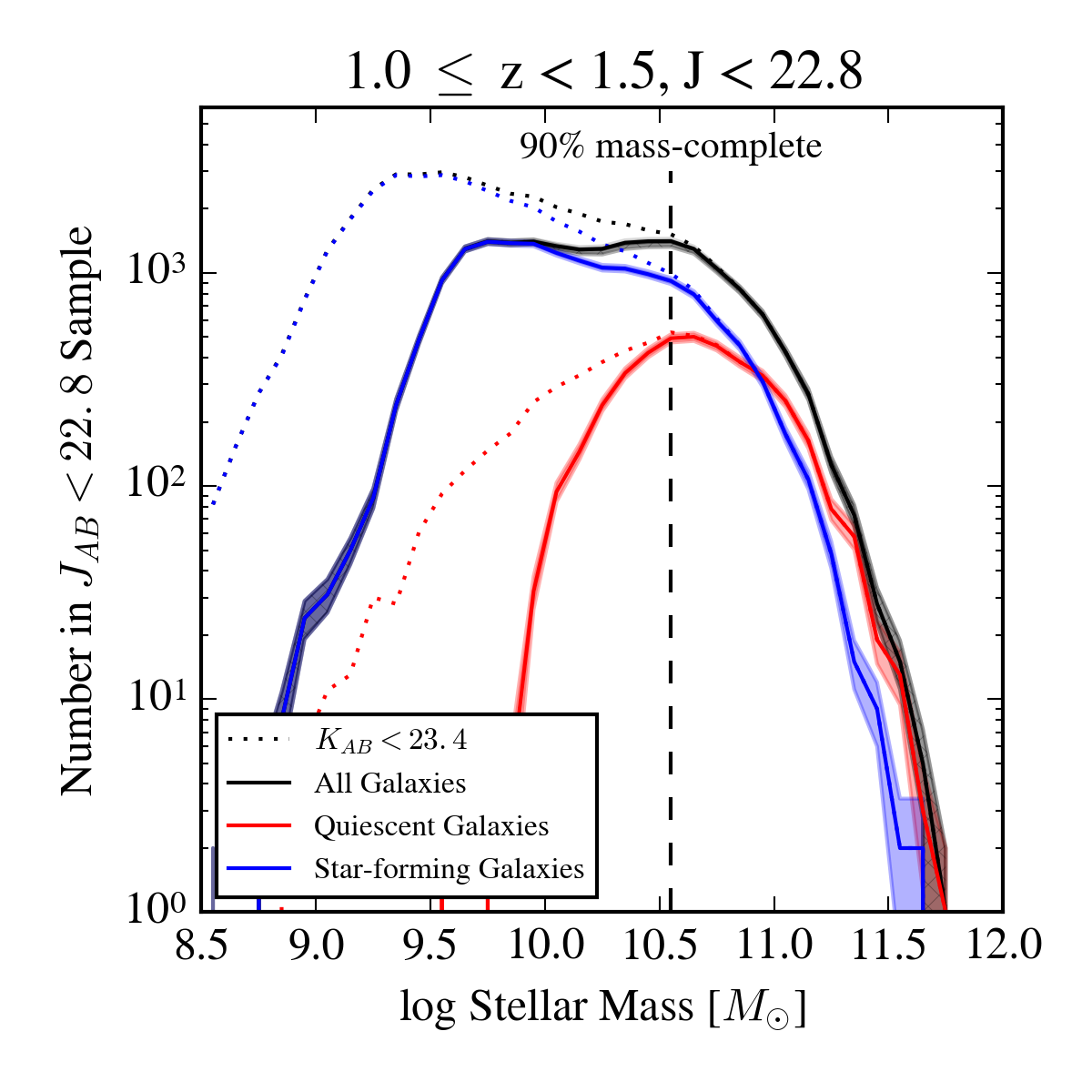}
\vskip -0mm
}
\figcaption[]{
\small
Distribution of galaxy mass for the $J$-band magnitude limit in COSMOS at $1{<}z{<}1.5$ (quiescent in solid red; star-forming in solid blue) relative to a $K-$band selected sample (dotted). Employing a $J-$band selection results in a 90\% mass-complete sample of galaxies of all colors to $M^* \sim 3 \times 10^{10}$~\msun, and a representative sample of both active and quiescent galaxies to below $L^* (M^* \sim 10^{10}$~\msun). 
\label{fig:massredshift}}
\end{figure}

\subsection{Continuum-selected galaxies ($z{<}2$)}
\label{sec:lowz}

The ``Main" $z{<}2$ sample is designed to map the large-scale structure, with galaxies observed for a total of 2 hours. We will also select a ``Deep" subsample, which will be observed for 12 hours in total. The selection and main science goals of these samples are described below.

{\it Main sample}: We will select targets with $0.7 {<} z {<} 1.7$ using a photometric redshift (photo-$z$)
selection (\S \ref{sec:photoz}), with a magnitude limit of $y=22.5$ over the full redshift range and $J=22.8$ mag for $z > 1$ galaxies (see Table \ref{tbl:target}). These two magnitude limits provide roughly constant stellar mass sensitivity down to $M_* \approx 3 \times 10^{10}$~\msun\ (90\% completeness). Our $y$- or $J$-selection falls redward of the 4000\AA\ break over our full redshift range, and thus is sensitive to stellar mass as well as star formation (see example spectra in Figure \ref{fig:snrspectra}, right).
Thus, we will chart the evolution of typical galaxies over this entire epoch, with 326k FH. 

{\it Deep sample}: We will spend an additional 144k FH on a deep spectroscopic component, targeting galaxies chosen from among the primary sample with measured redshifts.  We will study 14,000 galaxies with $J{<}22.5$ spread throughout the survey areas, split evenly between quiescent and star-forming galaxies (and thus not mass-representative).

\subsubsection{Area, Sampling, Depth} 

Our area is set by our desire to sample cluster-mass halos. In 11 PFS pointings (corresponding to $\sim$ 12.3 deg$^2$ after accounting for overlapping fields) we will cover a comoving volume of $\sim 0.1$~Gpc$^3$. We can recover the galaxy number density in $\delta M_{\ast}=0.5$ dex and $\delta z = 0.4$ bins to better than 0.03 dex for stellar mass ${<}10^{11}~M_{\odot}$ \citep[e.g.,][]{moster:2011}, such that we will not be cosmic-variance limited.
In the deliverables section (\S \ref{sec:deliverable}), we show that this area includes $\sim 35$ halos with $M_h>10^{14}$~\msun.

As described in detail in \S \ref{sec:deliverlss}, our high sampling (with mean inter-galaxy spacing between 7 and 10 Mpc) will allow us to measure clustering at scales of $0.1~h^{-1}\mathrm{comoving~Mpc}$, reaching the virial radius of $\sim 10^{12}$~\msun\ halos (Figure \ref{fig:massredshift}, right). \citet{Kraljic2017} and \citet{Malavasi2017} have reconstructed the cosmic web in lower redshift surveys with a similar range of inter-galaxy spacings.

The depth of the Main sample is chosen to achieve redshift success for the majority of targets (see \S \ref{sec:deliverable}). The Deep sample will require 12 hour integrations. With this depth, we will be able to measure reliable stellar velocity dispersions and star formation histories in the spectra of individual galaxies, and abundance ratios from the stellar continuum in stacks of 20-30 objects (\S \ref{sec:deliverable}). We will also measure weak emission lines such as [Ne III] and C III] in individual galaxies extending to $1\sigma$ below the star forming main sequence, allowing us to determine ionization parameter and gas-phase enrichment and abundance ratios. We will use measurements of the full suite of emission lines in the spectra of individual Deep galaxies to calibrate the use of the strong lines ([O II], [O III], H$\beta$) that will be commonly detected in the Main sample spectra. We can then stack individual Main sample galaxies in bins of stellar mass and star formation rate to determine nebular properties as a function of stellar mass and star formation rate \citep[e.g.,][Figure \ref{fig:massredshift}]{Strom:2022}.  Our sample will be large enough to identify 20-30 galaxies in each of 10 redshift bins ($\delta z=0.1$), 3-5 stellar mass bins and four bins in environment, as quantified by the distance to the nearest filament (\S \ref{sec:deliverlss}). This sampling therefore allows us to study trends in each of these quantities separately. 

In summary, our Main sample will comprise $\sim 230,000$ restframe-optical continuum-selected galaxies down to $M^* \approx 3 \times 10^{10}$~\msun\ at $0.7{<}z{<}1.7$, and integration times of 2 hours. It will cost 550k FH. This will be complemented by Deep spectra for 14,000 galaxies with integration times of 12 hours. Finally, we will supplement this with spectra of 8000 AGN selected using a variety of multi-wavelength techniques as described in the following subsection.

\begin{deluxetable*}{llcccccc}
\tabletypesize{\footnotesize}
\tablecaption{Full Target Table}
\tablehead{
\colhead{Redshift}&
\colhead{Selection}&
\colhead{Exp. Time}&
\colhead{Targets}&
\colhead{Sampling}&
\colhead{\# of}&
\colhead{Fiber} \vspace{-0.3cm}\\
\colhead{range}&&\colhead{(hrs)}&\colhead{PFS FOV}&\colhead{rate (\%)}&\colhead{spectra ($10^3$)}&\colhead{khrs}
} 
\startdata
\hline
Continuum-selected &&&&&\\
\hline
$\lesssim1$ & $i{<}23$ & 2 & 6100 & 40 & 24 & 48 \\
$0.7 - 1$ & $y{<}22.5$ + $z_{\rm ph}$ & 2  & 11900 & 50 & 58 & 116 \\
$1 - 2$ & $y>22.5+z_{\rm ph}$ & 2 & 11800 & 70 & 81 & 162 \\
&+ $J{<}22.8$&&&&&\\
$0.7 - 2$ & $y{<}22.8$ + $z_{\rm ph}$ & 12 & 1220 &...& 12 & 144 \\
\hline
Tomography &&&&&\\
\hline
$2.1 - 2.5$ & $y{<}24.3$ + $z_{\rm ph}$ & 6 & 8300 & 22 & 18 & 108 \\
$2.5 - 3.5$ & $y{<}24.3$ + $z_{\rm ph}$ & 6 & 770 & 90 & 6.8 & 41 \\
&+ $g{<}24.2$&&&&&\\
$2.5 - 3.5$ & $y{<}24.3$ + $z_{\rm ph}$ & 12 & 1800 & 65 & 11.5 & 138 \\
&+$24.2 {<} g {<} 24.7$&&&&&\\
\hline
LAE &&&&&\\
\hline
$3.5 - 7$ & $y{<}24.5$ + $z_{\rm ph}$ & 6 & 2500 & 74 & 18 & 108 \\
2.2 & log L$_{Ly\alpha}>42.5$ & 3 & 770 & 80 & 6.0 & 18 \\
5.7, 6.6 & log L$_{Ly\alpha}>42.7$ & 6 & 470 & 80 & 3.7 & 22 \\
5.7, 6.6 & log L$_{Ly\alpha}$=42.5-42.7 & 12  & 290 & 80 & 2.3 & 28 
\enddata
\tablecomments{Details of target selection. The first four entries summarize the continuum-selected ($0.7 {<} z {<}$ 1.7) sample, including the Main (2 hr) and Deep (12 hr) components. The next three rows outline details of the IGM tomography sample, and the final four rows focus on the very high redshift drop-outs and LAEs. We include the color selection, exposure times, number of targets in the 1.3 deg$^2$ field of view, the fraction of that population that we aim to target, the resulting total number of spectra and cost to the survey in fiber hours.}
\label{tbl:target}
\end{deluxetable*}

\subsection{Active Galactic Nuclei}

The discovery of a tight correlation between supermassive black hole (SMBH) masses ($M_{\rm BH}$) and the host galaxy properties, e.g., bulge masses, has transformed our understanding 
on the role of SMBHs in the process of galaxy evolution
\citep[e.g.,][]{2013ARA&A..51..511K}.
While this correlation may be due partly to frequent galaxy mergers in the hierarchical structure formation, recent 
studies indicate that baryon physics within individual galaxies or dark matter halos plays a fundamental role \citep[e.g.,][]{2014ARA&A..52..589H}. 
In particular, energy feedback from active galactic nuclei (AGNs) may have a significant impact on the evolution of galaxies. 
It is critical to know when, where, and how SMBHs have formed, evolved, and finally settled in the present-day universe, and what is the consequence 
of the AGN activity which accompanies SMBH growth
in the heart of galaxies.

The intrinsic SED varies from one AGN to another, depending on $M_{\rm BH}$, mass accretion efficiency (Eddington ratio $\lambda_{\rm Edd}$), and other factors.
Dust extinction further adds to the variety of observed SEDs.
Indeed, AGNs selected at different wavelengths are known to possess systematically different properties \citep[e.g.,][]{2009ApJ...696..891H}.
In order to obtain a full census of SMBHs  and their impact on the host galaxies, it is therefore imperative to observe various sub-populations of AGNs
selected in multiple complementary ways.
The PFS-GE sample described above and the [\ion{O}{2}] emitters targeted in the PFS-Cosmology survey will include numerous obscured AGNs, 
which we can identify with strong emission from the narrow line region, e.g.,
[\ion{O}{3}] $\lambda$4959, 5007 and [\ion{N}{2}] $\lambda$6548, 6583 lines \citep[e.g.,][]{1981PASP...93....5B}.
This section describes a strategy to target other classes of AGNs, exploiting existing multi-wavelength imaging data.

\subsubsection{AGN selection \label{sec:design}}

We select AGN targets from both the rich multi-wavelength data available in the 12.3-deg$^2$ Galaxy Evolution fields (\S \ref{sec:photometry}), from X-ray to radio frequencies, and the wide $\sim$1000-deg$^2$ area of the PFS Cosmology field to look for more luminous and rare AGNs. All the targets are selected from accurate HSC photometry. The targeting strategy of each sub-population is detailed below, and is summarized in Table \ref{tbl:agntargets}. A total of 53k FHs are planned for AGN science. We note that the AGN candidates presented here may partially overlap with galaxies meeting the selection criteria of the galaxy evolution (this document) or Cosmology targeting, so the fiber numbers listed in the table may not simply add to the numbers planned in the two main surveys.

A major part of our sample consists of candidate broad-line (BL) AGNs
across the whole range of redshifts observable with the HSC imaging, out to $z > 6$.
The target selection rests on all the available color information in the CFHT $u$ through {\it Spitzer} 4.5-$\mu$m bands.
We will establish a statistically robust sample of 3,000 BL AGNs at $0.5 {<} z {<} 4$, 
and target all 15,000 quasar candidates at $z > 4$ that were imaged by HSC and are feasible to detect with PFS. 
We will put high priority on the targets within 1 arcmin of bright SDSS quasars at $z \sim 3$ in the GE field,
which will allow us to probe the CGM/IGM around the targets via absorption of the background quasar spectra
\citep[e.g.,][]{2013ApJ...776..136P} in an unprecedented number of sightlines.

We will also observe 2,000 X-ray sources, in order to probe the obscured and possibly the most active phase of SMBH 
and galaxy assembly history \citep[e.g.,][]{2018MNRAS.478.3056B}.
The target selection rests on public {\it XMM-Newton} and {\it Chandra} deep images as well as the all-sky {\it eROSITA} survey data,
putting higher priority on objects with $z_{\rm phot} > 1$.
We will further target dusty AGNs; 1,500 WISE sources with $f_{22{\mu}m} > 3$ mJy in the Cosmology field lacking SDSS spectra,
and 1,000  JCMT/SCUBA-2 sub-mm sources with $f_{870{\mu}m} > 5$ mJy that have ALMA counterparts in the HSC Deep fields 
\citep{2018ApJ...860..161S, 2020MNRAS.495.3409S}, both expected to include a significant fraction of AGNs
\citep[e.g.,][]{2006ApJ...644..143B, 2009ApJ...703.1778S,2015PASJ...67...86T, 2017ApJ...840...78D}.
We also aim to capture SMBHs at both the beginning and end of their most active phase. 
Once they have become sufficiently massive, SMBHs are often observed to produce large-scale ratio jets \citep[e.g.,][]{2004MNRAS.353L..45M} and may be found
in over-dense regions \citep[e.g.,][]{2008A&ARv..15...67M}.
We will observe 2,000 candidate radio AGNs obtained by cross-matching the HSC and the FIRST survey catalogs. At the other extreme, black holes before experiencing significant growth may be identified as intermediate-mass black holes (IMBHs) 
with $M_{\rm BH} \lesssim 10^{4-5} M_\odot$ in the local universe \citep[e.g.,][]{2020ARA&A..58..257G}. We will look for IMBHs by targeting 300 candidate low-luminosity AGNs, identified via time variability in HSC Deep fields \citep{Kimura:2020}.

Finally, we aim to observe 1,100 lensed quasar and galaxy candidates in the PFS Cosmology field.
Confirmed systems will offer a wide range of astrophysical and cosmological applications \citep[e.g.,][]{2006AJ....132..999O}, including measurements of the mass 
structures of the foreground lenses, mapping the host galaxies of the magnified background quasars, and determination of the Hubble constant.
The candidates were selected from the HSC imaging, and are expected to include $\sim$100 lensed quasars as well as
$\sim$900 lensed galaxies 
\citep[][]{2018PASJ...70S..29S, 2020MNRAS.495.1291J}, including $\sim$300 galaxy-scale and $\sim$600 group-to-cluster scale lenses.

\begin{deluxetable*}{llrrrrr}[h]
\tabletypesize{\footnotesize}
\tablecaption{AGN target selection and observing strategy}
\tablehead{
\colhead{Targets} &
\colhead{Selection} &
\colhead{$N_{\rm AGN}^{\rm total}$} &
\colhead{$N_{\rm AGN}$} &
\colhead{$N_{\rm fiber}$} &
\colhead{$T_{\rm exp}$} &
\colhead{$N_{\rm fiber} T_{\rm exp}$} 
} 
\startdata
Galaxy Evolution &  &  &  &  &  & \\
\hline
BL candidates ($z {<} 4$)    & CFHT $u$ -- {\it Spitzer} colors        & 5,700  & 3,000  &   6,000 (0.5) & 1 -- 4 & 15,000\\
BL candidates ($z > 4$)    & HSC -- {\it Spitzer} colors                & 500      &    500 &   1,000 (0.5) & 4 -- 5 & 4,500\\
X-ray sources                   & {\it Chandra}, {\it XMM-Newton}       & 10,000   &  2,000 &   2,000 (1.0) & 4 -- 5 & 9,000\\
Sub-mm galaxies              & SCUBA-2 w/ ALMA counterparts   & 300 & 300 &  1,000 (0.3) & 5 & 5,000\\
Radio galaxies                  & FIRST                                              & 200 & 200  & 300 (0.7) & 3 & 900 \\
IMBH candidates              & HSC flux variability                           &          30 &      30  &  300 (0.1)  & 2 & 600\\
\hline
Total                                     &                                          &                        & 6,030          & 10,600      &                         & 35,000\\
\hline
\hline
Cosmology &  &  &  &  &  & \\
\tableline
BL candidates ($z > 4$)         & HSC colors             & 15,000   & 15,000   & 30,000 (0.5) & 0.5 & 15,000  \\
X-ray sources                        & {\it eROSITA}          & 1,700   &  1,700       &  1,700 (1.0) &  0.5 & 850\\
Mid-IR sources                      & WISE 22 $\mu$m   & 1,000     & 1,000     & 1,500 (0.7)   & 0.5 & 750  \\
Radio galaxies                       & FIRST                     & 20,000 & 1,500     & 1,700 (0.9)   & 0.5 & 850\\
Lensed quasar candidates     & HSC shapes            & 100       & 100        &  1,100 (0.1) & 0.5 & 550\\
\hline
Total                                     &                                &              & 19,300   & 36,000        &       & 18,000\\
\enddata
\tablecomments{Columns are: Target; Selection method; Total number of AGNs expected over the entire survey field; Number of AGNs we aim to observe; 
Number of requested fibers (the number in parenthesis represents the expected success rate of AGN identification, i.e., $N_{\rm AGN}$/$N_{\rm fiber}$); Exposure time (hr); Requested fiber hours.}
\label{tbl:agntargets}
\end{deluxetable*}

\section{Photometry for target selection and parameter estimation}
\label{sec:photometry}

Each component of the survey relies on multi-wavelength imaging for selection. In this section, we review the available photometry and depths, along with the photometric redshift precision that can be achieved with this photometry. The basis of the survey are the HSC Deep fields, and specifically the $\sim 12.3$~deg$^2$ with complementary NIR imaging (required to selected the Continuum sample, whose 4000\AA\ or Balmer breaks falls in the $J-$band at $z \approx 2$). The HSC-Deep fields that already contain $J-$band photometry down to $\sim 23.7$ mag or better are XMM-LSS, DEEP2-3, and E-COSMOS. These fields already have {\sl Spitzer} 3.6 and 4.5$\micron$ IRAC data to a similar depths of $\sim 23.5$  mag, which is useful in constraining the stellar masses of our galaxies. The CFHT Large Area $U-$band Deep Survey (CLAUDS) project has obtained 20 deg$^2$ of $u-$band data to $u = 27$~mag ($5\sigma)$ with sub-arcsec seeing in the HSC Deep Fields which, when combined with the $JK$+IRAC CH1+2 photometry, will allow excellent determination of photo-$z$ (\S \ref{sec:photoz}). These fields also have a range of other multiwavelength data, including X-rays, far-IR/sub-mm, and radio, which all contribute to constraining the accretion activity of the central supermassive black holes across different levels of dust obscuration and accretion rates. 

\begin{figure*}
\begin{center}
\includegraphics[trim=1cm 12cm 5cm 1.5cm,clip,angle=0,width=0.4\textwidth]{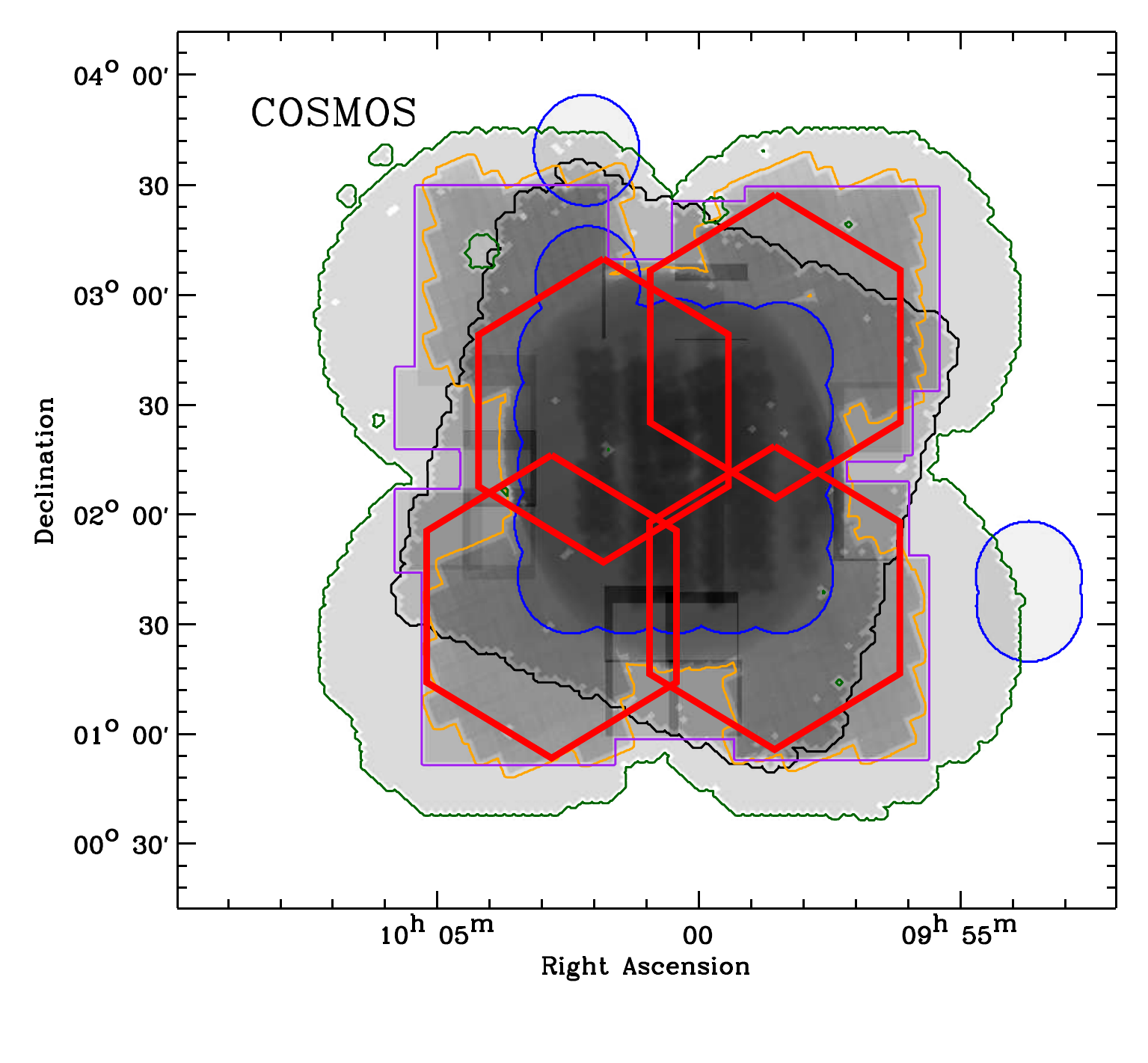}
\includegraphics[trim=1cm 12cm 5cm 1.5cm,clip,angle=0,width=0.4\textwidth]{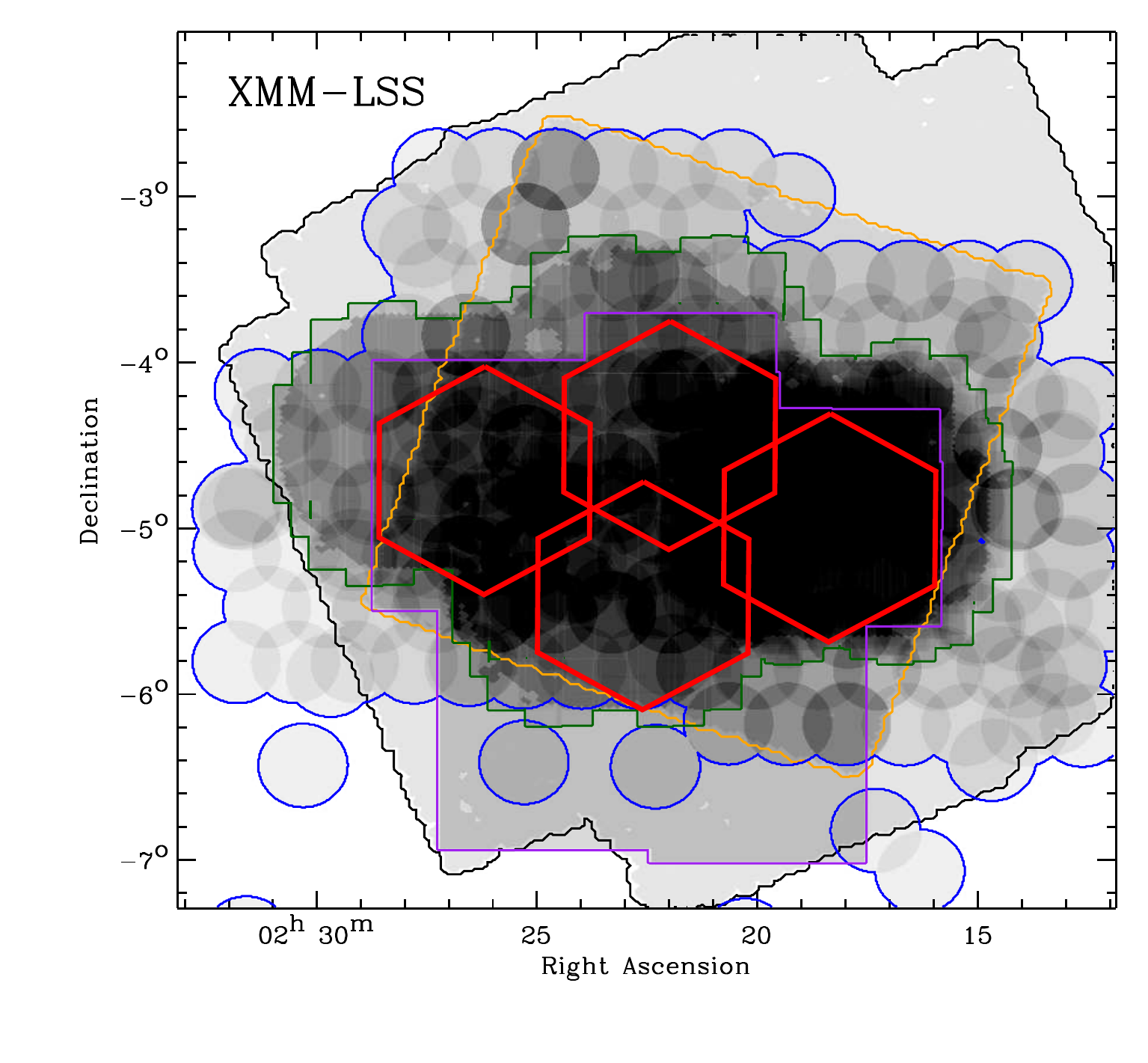}
\includegraphics[trim=1cm 12cm 5cm 1cm,clip,angle=0,width=0.4\textwidth]{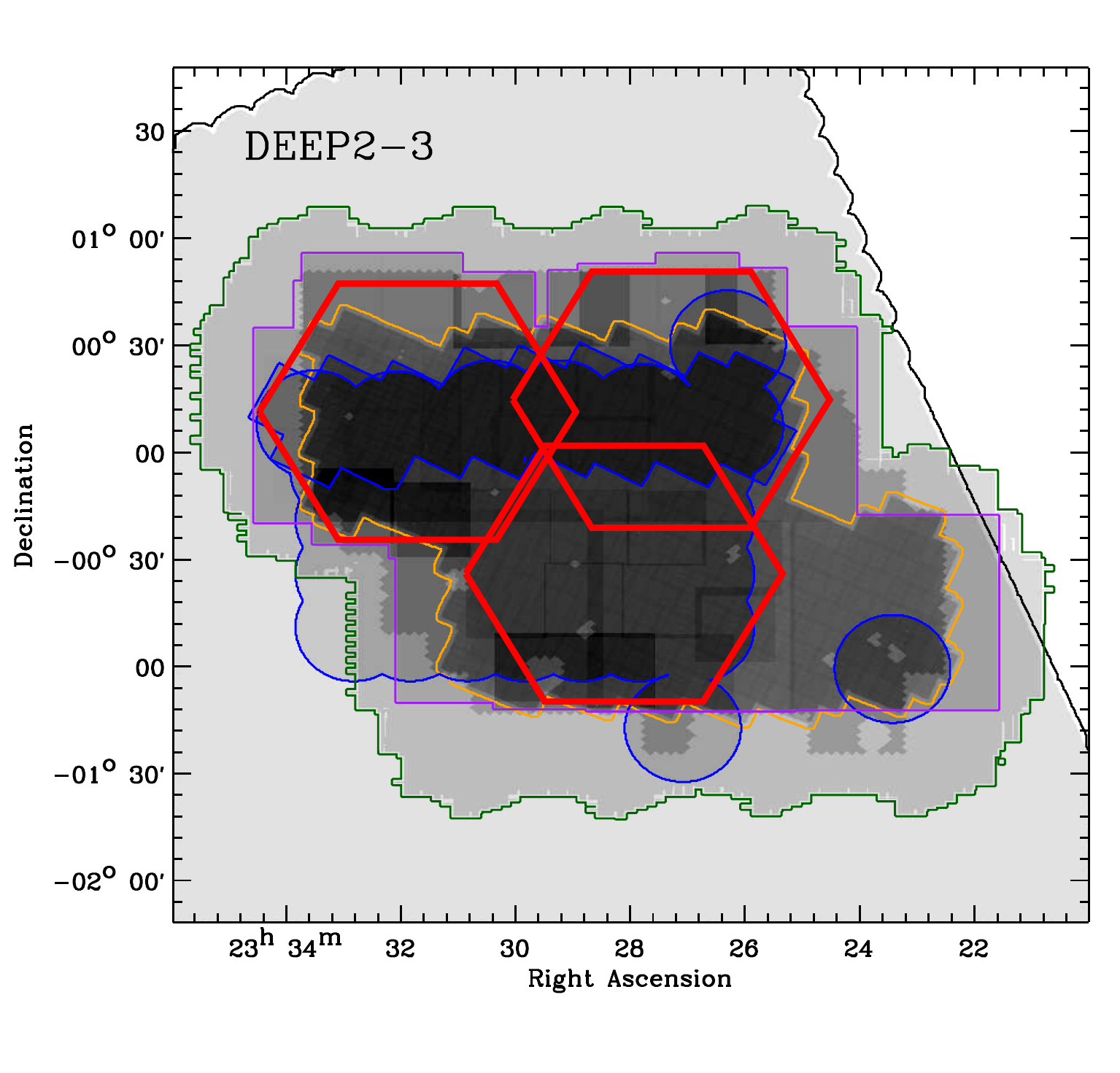}
\includegraphics[trim=-1cm -2cm 0cm 0cm,clip,angle=0,width=0.38\textwidth]{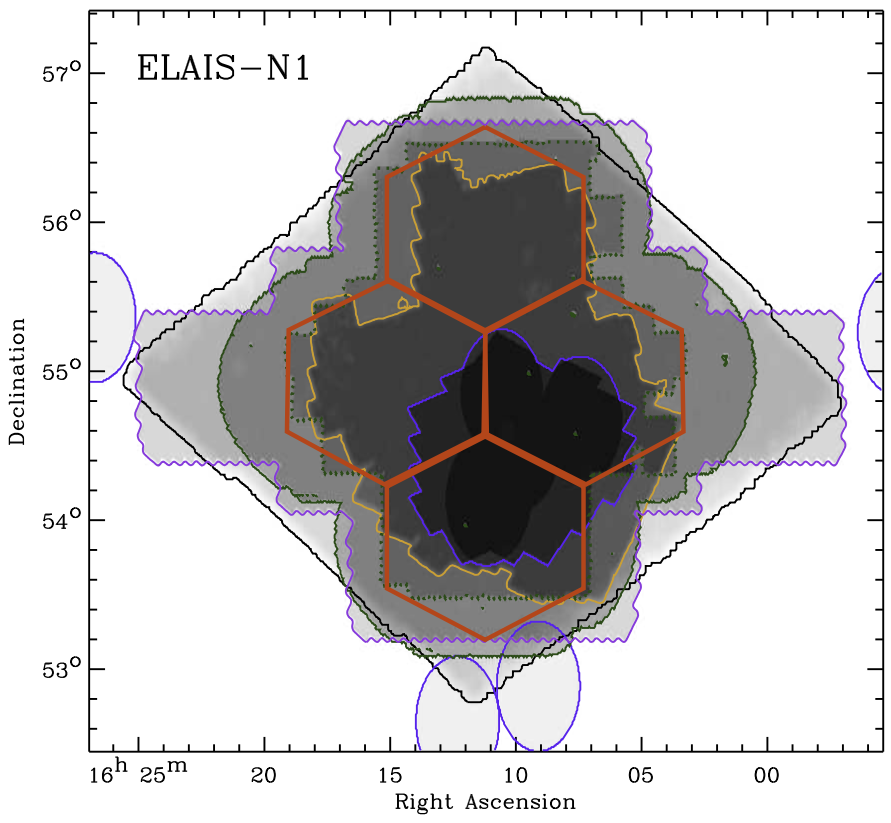}
\figcaption[]{PFS pointings comprising our survey (red) superimposed on multi-wavelength data available in each field, illustrating how we have selected optimized pointings to maximize multi-wavelength overlap. Different colors identify different data. Green: HSC+CLAUDS; Orange: Spitzer; Purple: NIR (VISTA+WIRCAM+UKIDSS); Blue:X-ray (Chandra+XMM); Black: Herschel. Grey: the coverage for each survey band, such that the darkest regions have the most overlapping multi-wavelength coverage.
Red: PFS. Note that EN1 is not part of the PFS survey area, but may be used for target exploration.}
\label{fig:field_choice}
\end{center}
\end{figure*}


We summarize the 12.3 deg$^2$ area with the PFS pointings overlaid in Figure \ref{fig:field_choice}. We chose the pointings to optimize overlap with multi-wavelength data-sets (summarized below) and in particular overlap with $J-$band imaging, which is crucial for our success.

\subsection{HSC Deep}

The four fields of the HSC-SSP \citep{Aihara2018} Deep layer cover $32$~deg$^2$ with five broad-band filters ($grizy$) and a suite of narrow-band filters plus a wealth of ancillary data,  making them ideal deep fields for studying galaxy evolution over a broad range of galaxy mass, redshift and environment. 
These fields (XMM-LSS, E-COSMOS, ELAIS-N1, and DEEP2-3) reach a depth of $\sim26.5$ mag ($5\sigma$ point sources), and contain
the HSC UltraDeep layer with two separate 1.75 square degree fields: COSMOS and the Subaru/XMM-Newton Deep Survey (SXDS) that reach a depth of $i\sim$ 28  mag. It is also worth noting the rich array of science enabled by the HSC Deep data has many synergies with the science goals outlined here, including weak lensing \citep[e.g.,][]{Huang2020,Luo:2022}.

Deep HSC narrow-band (NB) imaging provides large samples of line-emitting galaxies up to $z\sim7$ \citep[e.g.,][]{shibuya2018a}. The  HSC-SSP includes NB images centered at 387, 816 and 921 nm over most of the HSC Deep area that are effective at detecting galaxies with strong emission lines \citep{Hayashi2020} including H$\alpha$, [OIII]$\lambda$5007 ($z=0.840, 0.939$), [OII]$\lambda3727$ ($z=0.92, 1.19, 1.47, 1.70$), and Ly$\alpha$ ($z=2.18, 5.73 ,6.58$). The program ``Cosmic HydrOgen Reionization Unveiled with Subaru'' (CHORUS; \citealt{Inoue:2020}) supplements these with additional filters at 387, 527, 718 and 973 nm in the HSC UD-COSMOS field at 5$\sigma$ depths $\sim$26 (see Table 2 in \citealt{Inoue:2020}). Of interest for the PFS survey, CHORUS extends the samples of LAEs to redshifts of 3.33, 4.90 and 6.99.

\subsection{Near-IR Imaging}

At the beginning of the HSC survey
(ca.~2014), the majority of the HSC deep fields lacked NIR data of
comparable depth to the optical data from the HSC. 
To remedy this, a team from the University of Arizona started the Dunes$^2$ survey
(PIs: E.~Egami and X.~Fan) in 2015, using UKIRT to survey the flanking
fields of the E-COSMOS, DEEP2-3, and part of the ELAIS-N1,
to $J\approx 23.5$ and $K\approx 23.2$ (E.~Egami et al.~in prep.).

Even with this effort, both the depth and area covered are still
insufficient for carrying out a robust target selection of the PFS GE
$z=0.7-1.7$ component.  Another survey called DeepCos, using both
UKIRT and CFHT, aims to bring the depth uniformly to $J=23.7$
over 12.3\,deg$^2$ 
(in E-COSMOS, DEEP2-3, and XMM-LSS) and was initiated and coordinated by Y.-T.~Lin (ASIAA).\footnote{With
significant contributions from W.-H.~Wang (ASIAA), M.~Ouchi (The
University of Tokyo/NAOJ), Z.~Cai and C.~Li (Tsinghua University), and
A.~Goulding (Princeton).}  Upon the completion of this survey (in 2023) we will have sufficient NIR coverage for targeting.

\subsection{Spitzer IRAC}

{\sl Spitzer} IRAC 3.6\,$\mu$m and 4.5\,$\mu$m coverage with good uniformity and 5$\sigma$ depth of 23.7 (23.3) magnitude for 3.6\,$\mu$m (4.5\,$\mu$m) exists for the vast majority of the HSC-Deep fields. This coverage is primarily thanks to the SERVS \citep{Mauduit2012} and {\sl Spitzer} DeepDrill surveys \citep{Lacy2021} as well as a dedicated $\approx$500hr {\sl Spitzer} GO14 program ({\it SHIRAZ}; PI: A.~Sajina) to complete the IRAC coverage of the HSC-Deep fields. This program is completed, the data are reduced and mosaics are nearing publication (Annunziatella et al.~2022, AJ submitted). Specifically, we have re-reduced all archival data for consistency and our mosaics for E-COSMOS, ELAIS-N1 and DEEP2-3 incorporate both the new and archival data for a total of $\approx$17\,deg$^2$. 

IRAC data are critical for extending the wavelength coverage and thus improving photo-$z$ and stellar population parameter estimates \citep[see e.g.,][\S \ref{sec:photoz}]{Muzzin2009}, and by extension, our target selection. We note that the above depth is measured from the IRAC catalog. However, work is in progress on multi-band forced-photometry using shorter wavelength and higher-resolution data (such as HSC $z$ or $y$) as priors. Based on our experience \citep[e.g.][]{Nyland2017}, using such priors allow us to reach at least 0.5 magnitudes deeper (e.g.,  $\approx$24.2 magnitude at 3.6\,$\mu$m). As described by \citet{Nyland2017}, this is because we can mitigate blends and because the prior positional knowledge allows us to glean spectral information from the IRAC data even when the detection is  $<5 \sigma$ without the prior positional information. At these depths, we should be able to reach galaxies with stellar mass $\approx$10$^9$M$_{\odot}$ at $z=1.0$ and $\approx$10$^{11}$M$_{\odot}$ at $z=5$. 


\subsection{Photometric redshifts and calibration}
\label{sec:photoz}

The PFS Galaxy Evolution targeting will rely on CLAUDS+HSC+NIR photometry. The CLAUDS team has performed uniform $ugrizyJHK$ photometry in the XMM-LSS and COSMOS fields, using near-infrared data from the Visible and Infrared Survey Telescope for Astronomy (VISTA) Deep Extragalactic Observations (VIDEO) survey \citep{Jarvis2013} in XMM-LSS and from UltraVISTA \citep{mccracken2012} in COSMOS. Using the CLAUDS+HSC+NIR data at the nominal depth in the COSMOS and XMM-LSS fields, we directly test the photo-$z$ accuracy. We compare our preliminary HSC photo-$z$ with objects that have spectroscopic redshifts and obtain encouraging results. We achieve excellent redshift success, with $\sigma_{\Delta z/(1+z)}{<}0.03$ [0.04] with NIR [no NIR] and low contamination fraction 4\% [5\%] below $z{<}2.5$, where we will use photo-$z$ selection. These tests are explored in detail in Deprez et al.\ (2022, submitted).

\subsection{Science Commissioning}

In the first year of the SSP survey, we will spend $\sim 10$ nights on science commissioning. Specifically, we will relax our color cuts and take a representative set of spectra down to our magnitude limit of $J=22.8$~mag, in order to (a) calibrate our photometric redshifts with a more representative sample in color space and then (b) refine our color selection accordingly. We will do this testing in the fourth HSC-Deep field, ELAIS-N1 (Figure \ref{fig:field_choice}) and then use the results from this field to define the precise definition of the main survey design. These calculations will be done in combination with our mock universe simulation (\S \ref{sec:mock}).

\begin{figure*}
\vbox{ 
\vskip -1mm
\hskip +15mm
\includegraphics[angle=0,width=0.85\textwidth]{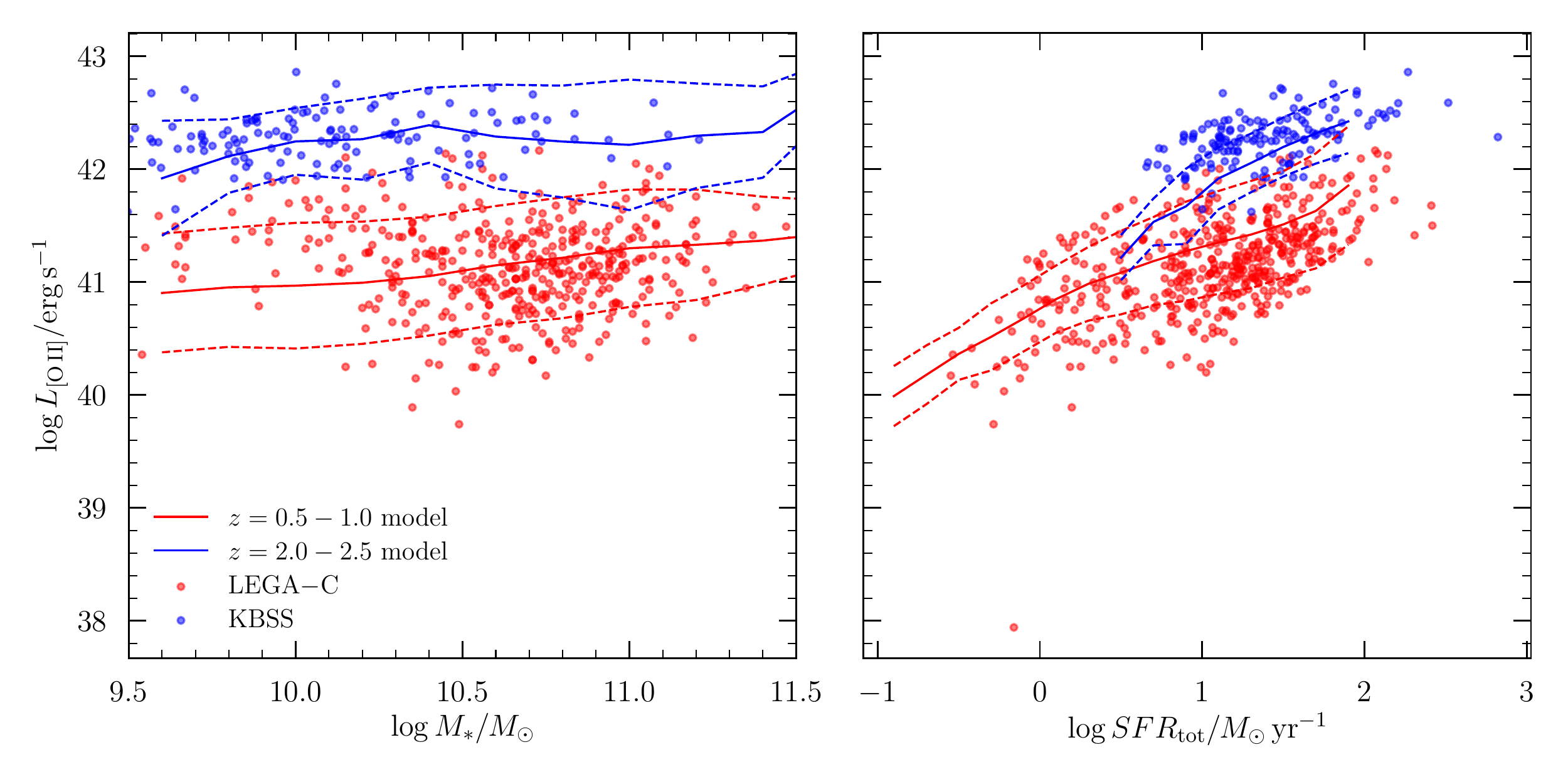}
}
\vskip -0mm
\figcaption[]{
\small
Comparison of the emission-line strengths in simulated PFS galaxy spectra and real galaxies observed at $z=0.5-1$ and $z=2.0-2.5$. This comparison ensures that our simulated spectra exhibit representative emission line strength distributions. {\it Left}: Simulated [O\,II] emission line luminosity versus stellar mass with the mock galaxy spectra at $z=0.5-1.0$ and $2.0-2.5$ shown in blue and red respectively. The solid line marks the running median and dashed line marks $1\sigma$ scatter as a function of stellar mass.
{\it Right}: Same as the left panel but with [O\,II] luminosity versus SFR. In both panels, galaxies at $z=0.5-1.0$ from LEGA-C and at $z=2.0-2.5$ from KBSS-MOSFIRE are shown as red and blue points respectively.
\label{fig:legactest}}
\end{figure*}

\section{Mock PFS Surveys}
\label{sec:mock}

In order to perform end-to-end simulations of the survey, we need a mock galaxy catalog that (a) has realistic spectral features and realistic PFS-like noise on a per-galaxy basis and (b) that tells us about the relative positions of galaxies, both physically and in projection. To accomplish this goal, we build a mock catalog as described in the following two sections. 

We generated a mock catalog of galaxies of similar volume to the planned survey using data from the UniverseMachine DR1 \citep{Behroozi19}. The UniverseMachine is an empirical model which has been calibrated to statistics of galaxy populations over $0{<}z{<}10$. This model places a galaxy at the center of each halo from the Bolshoi-Planck cosmological $N$-body simulation \citep{Klypin:2016}, and assigns SFRs to track the accumulation of stellar mass following the dark matter assembly for each galaxy. From the UniverseMachine models, we constructed light cones over a randomized viewing angle and origin inside of the periodic cube, with volumes matched to the PFS survey specification \citep[see details in ][]{pearl:22}. 

At $z>2$, we use abundance matching to ensure we match the number density of sources. At lower redshifts, $z{<}2$, we need to adapt the mapping between mass and light in the default UniverseMachine catalogs to match our sample properties. Specifically, we assigned fluxes in several filters to each mock galaxy in the light cone at $z{<}2$, calibrated to UltraVISTA photometry in the COSMOS field \citep{muzzin:2013}. This photometry is assigned onto the physical properties as a multivariate mapping from the specific star formation rate to the mass-to-light ratio in each observed-frame filter. To perform this mapping, we trained a random forest on UltraVISTA sources, using the SFR inferred from UV photometric bands and stellar mass derived from SED fitting. These SEDs were fit by the spectral fitting code Fitting and Assessment of Synthetic Templates \citep[FAST][]{Kriek2009}, using the \citet{chabrier2003} initial mass function, an exponentially declining star formation history, and the \citet{Bruzual2003} stellar library. For each galaxy in the sample, our mock survey contains a precise spectroscopic redshift, position on the sky, and multiwavelength photometry. Because of its empirical calibration, the mock provides excellent agreement with the color, redshift, and mass distributions of galaxies in the real $0.7{<}z{<}1.7$ Universe \citep{pearl:22}. This mock facilitates a number of targeting and fiber assignment tests as well as predictions for the survey performance.

\subsection{PFS Simulator}

Our model for the throughput of the spectrograph includes the quantum efficiency of the chips, light losses at the fiber due to seeing, losses in the fiber and fiber coupling, reflectivity of the Subaru primary mirror and transmission of the corrector, throughput of the camera and collimator (including vignetting) and dichroics as a function of wavelength, the grating transmission, and the atmospheric transparency. The details of the instrument throughput and the expected performance are summarized for the public\footnote{\url{https://pfs.ipmu.jp/research/parameters.html}}. The spectral simulator models the two-dimensional distribution of source and sky photons on the chip, then collapses along the spatial dimension to produce simulated one-dimensional spectra and the corresponding error arrays. The noise calculation includes shot noise from the source, sky emission lines, and sky continuum, as well as read noise, dark current, and thermal background. Furthermore, an additional noise of 1\% of the sky background is added as a possible systematic sky subtraction error. In the spectral simulation, we assume that the moon phase is dark, the seeing is FWHM=0.8 arcsec, and the effective radius of galaxies is 0.3 arcsec.

In order to guide the PFS survey design and ensure that the spectral quality is sufficient for our science goals, we produced a large mock catalog of galaxy spectra to run through the PFS spectral simulator.  As a starting point, we took the COSMOS+UltraVISTA photometric galaxy catalog \citep{muzzin:2013} described above and applied redshift and magnitude limits corresponding to the different survey components. We then adopted the best-fit star-formation history from COSMOS+UltraVISTA and produced a mock stellar spectrum for each galaxy using the model of \citet{Bruzual2003}.


While this produces a reasonable distribution of galaxy spectra tied to the photometric redshifts and magnitudes of real galaxies, we must also include emission lines and ISM absorption features. Furthermore, the spectral models have insufficient spectral resolution in the NUV and FUV. To overcome these limitations, we performed the following post-processing to produce the PFS mock spectra.
\begin{enumerate}
	\item In the near-UV, the simple stellar population models have low resolution so we take continuum-normalized NUV spectra from SDSS-BOSS stacks and insert them onto the low-resolution simple stellar population model continuum. This inserts realistic NUV ISM absorption and emission lines such as Mg~II, Fe~II, Fe~II*, etc. We created $6$ different NUV stacks using SDSS objects at different redshifts and with different SFRs and adopted the stack that best-matches the redshift and star formation rate of the corresponding COSMOS$+$UltraVISTA galaxies. 
	\item In the FUV, we follow the same approach except using stacks from the KBSS-MOSFIRE survey \citep{steidel14a} in mass bins so that real correlations between Ly$\alpha$ strength, ISM absorption strengths, and mass are present in our simulation. The KBSS-MOSFIRE galaxies have redshifts measured from non-resonant, rest-frame optical emission lines ensuring (a) that  redshift uncertainty does not degrade the effective resolution of the stacked spectra and (b) that the objects have realistic offsets between ISM absorption velocities ($v_{\rm out}$) and the galaxy systemic velocities.
	\item The strong optical emission line are set based on \citet{Valentino:2017}, which are calibrated at $z \approx 1.5$. In detail, the [O II] and H$\alpha$ emission are set based on the model SFR while the [O III] and other line ratios are set based on correlations with stellar mass. Dust extinction is included based on the SED and a nominal gas/stellar extinction ratio. We test the emission-line strengths as a function of mass and SFR at both $z \approx 0.8$ and $z\approx 2.2$ by comparing with observed scalings from LEGA-C and KBSS (Figure \ref{fig:legactest}) and find reasonable agreement.
	\item In addition, we perform an independent set of simulations with the optical line strengths set based on the UV SFR and low-$z$ metallicity-line ratio relations from \citet{Kewley:2002} and \citet{Maiolino:2008}. These mocks represent a worst-case scenario because the expected emission line strengths are lower when using the low-$z$ relations.
	\item For a small set of the galaxies, we also add higher ionization emission lines such as [Ne V] and higher [O\,III]/H$\beta$ ratios as expected from obscured AGN based on observed line strengths from \citep{yuan2016}.
	\item For galaxies at $z\gtrsim 2$, IGM absorption from the H\,I Ly$\alpha$ forest significantly impacts the FUV spectra of galaxies. To accurately include these Ly$\alpha$ forest lines, for each $z>2$ galaxy we identified a quasar with a similar redshift from the SQUAD database \citep{Murphy:2019} and multiplied the mock galaxy spectrum by the continuum normalized UVES spectrum of the quasar after degrading the UVES spectral resolution to match PFS.
	\item We create an additional set of LAE spectra specifically to explore recovery of LAE line shapes. We use numerical simulations of LAEs whose spectra with Ly$\alpha$ emission are determined by radiative transfer calculations that reproduce the HSC LAE samples in the partly neutral IGM \citep{inoue2018}.

\end{enumerate}

\section{Deliverables}
\label{sec:deliverable}

We summarize here our target fidelity in key observables. We start with redshifts, and we will demonstrate redshift fidelity of $\Delta z/(1+z){<}5 \times 10^{-4}$. We focus on the redshift fidelity of LAEs specifically. This level of redshift success is assumed in the recovery tests that follow. We show that with PFS data, we will quantify {\it stellar masses} to $\sim 0.2$ dex, {\it star formation rates} of a few $M_{\odot}$ per year,  {\it gas-phase metallicities} to ${<}0.25$~dex, and outflow velocities (in stacks) to a few percent.
For the Deep $0.7{<}z{<}1.7$ sample, we will determine {\it stellar and gas velocity dispersions} for the Deep galaxy sample to $\sim 15\%$ and {\it star formation histories} to 20\% for star-forming galaxies and even very low specific star-formation rates to within a factor of two. In the AGN samples, we expect to achieve {\it black hole masses and Eddington ratios} in broad-line AGN to 0.2 dex. 

\subsection{Redshift fidelity}
\label{sec:deliverablez}

\begin{figure}[]
    \centering
        \includegraphics[width=0.45\textwidth]{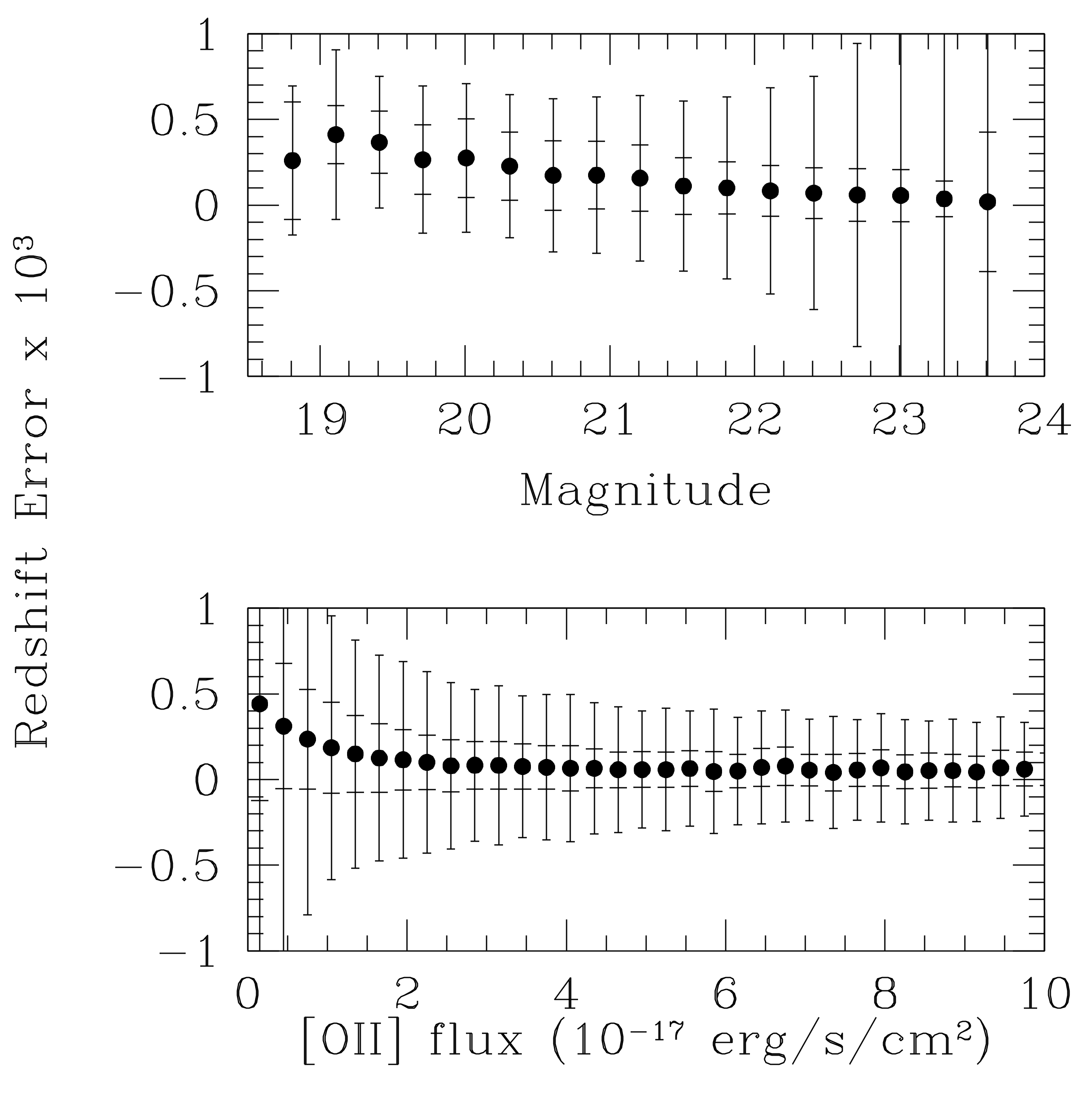}
        \caption{The median (points), interquartile range (smaller errorbars) and range including 90\% of the points (larger errorbars), of the difference between true and inferred redshift for a sample of simulated GE galaxies (\S \ref{sec:deliverablez}). Results are shown as a function both of galaxy magnitude and [OII] emission-line strength. The vast majority of galaxies have redshifts determined to within $|\Delta z| = 10^{-3}$.}
        \label{fig:mag_error}
\vskip -3mm
\end{figure}

The galaxy redshifts will be measured with a code called ``Algorithms for Massive Automatic Z Evaluation and Determination'' \citep[AMAZED;][]{Schmitt:2019}, which has been
used extensively for ground-based multi-object surveys such as the VIMOS VLT Deep Survey \citep{lefevre:2005}, and will be used to analyze data from the Euclid survey \citep{Laureijs:2011}.  AMAZED models galaxies as a superposition of continuum, emission lines, stellar and interstellar absorption lines, and including a model for the Ly$\alpha$ forest as a function of redshift.  These models are fit to the data by minimizing $\chi^2$, resulting in an estimate of the redshift and its uncertainty, as well as emission-line and absorption-line parameters.  We have tested this approach extensively on simulated PFS spectra and we have found that 90\% of the simulated galaxy sample with $z {<} 2.1$ have measured redshifts with $\Delta z/(1+z){<}5 \times 10^{-4}$, and 98\% have errors less than $10^{-3}$ (Figure~\ref{fig:mag_error}). 

\begin{figure*}
\begin{center}
\center{\includegraphics[trim=0.0cm 0cm 0cm 0cm, clip,angle=0,width=1.0 \textwidth]{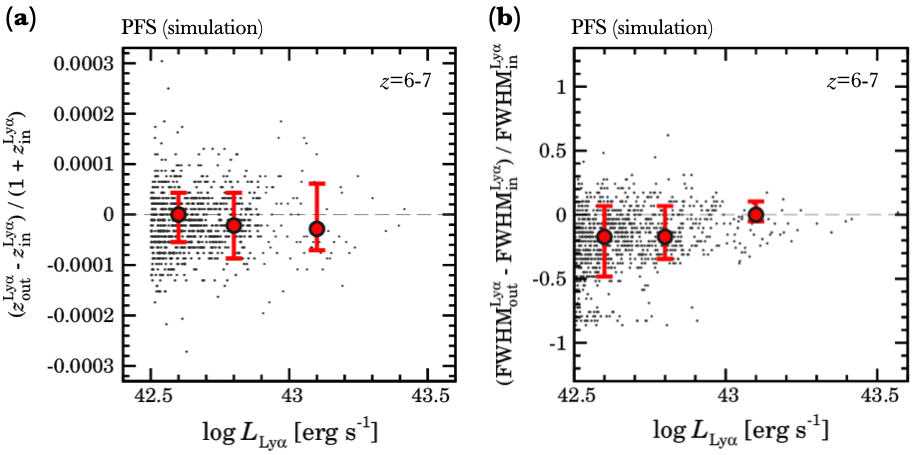}}
\figcaption[]{
(a) Simulation results with mock PFS LAE spectra to determine the accuracy of the LAE redshifts
as a function of Ly$\alpha$ luminosity ($\log L_{\rm Ly\alpha}$) for LAEs at $z=6-7$.
The LAE redshift determination accuracies are represented with
($z_{\rm out}^{\rm Ly\alpha} - z_{\rm in}^{\rm Ly\alpha})/(1  + z_{\rm in}^{\rm (Ly\alpha})
\equiv \delta z / (1+z_{\rm in}^{\rm Ly\alpha}$),
where $z_{\rm in}^{\rm Ly\alpha}$ ($z_{\rm out}^{\rm Ly\alpha}$) is the
redshift of the input spectra (the measurements with the mock PFS LAE spectra). The black circles are the individual LAEs, while the red circles and the errors
indicate the average values and the $1\sigma$ scatters.
(b) Same as (a), but for FWHM measurements. Here
the Ly$\alpha$ FWHM determination accuracies are similarly shown with ($FWHM_{\rm out}^{\rm Ly\alpha} - FWHM_{\rm in}^{\rm Ly\alpha})/FWHM_{\rm in}^{\rm Ly\alpha}
\equiv \delta FWHM / FWHM_{\rm in}^{\rm Ly\alpha}$,
where $FWHM_{\rm in}^{\rm Ly\alpha}$ ($FWHM_{\rm out}^{\rm Ly\alpha}$) is the
FWHM of the input spectra (the measurement with the mock PFS LAE spectra).
\label{fig:simulations_redshift_FWHM}}
\end{center}
\end{figure*}

We also investigate the accuracy of the Ly$\alpha$ redshift and FWHM determinations using the mocks described in \S \ref{sec:mock}. We find that the $1\sigma$ scatter of the redshift differences between the inputs and mock PFS LAE spectrum measurements $\delta z$ is within $\delta z =0.0003$ down to our luminosity limit of $\log L_{\rm Ly\alpha}/[{\rm erg\ s^{-1}}]=42.5$. The right panel of Figure \ref{fig:simulations_redshift_FWHM} shows the FWHM measurements of the mock LAE spectra. The FWHM differences between the inputs and mock PFS spectrum measurements $\delta FWHM = 40$ km s$^{-1}$. We assume a contamination rate of 10\% in the photometric LAE sample revealed by previous spectroscopic studies with Suprime-Cam and HSC \citep{ouchi2010,shibuya2018a}, and adopt a success rate of 90\% in deriving our PFS feasibility results. 

\subsection{Stellar Mass} 
\label{sec:delivermass}

\begin{figure*}
\hskip 15mm
\vbox{
\includegraphics[angle=0,width=0.75\textwidth]{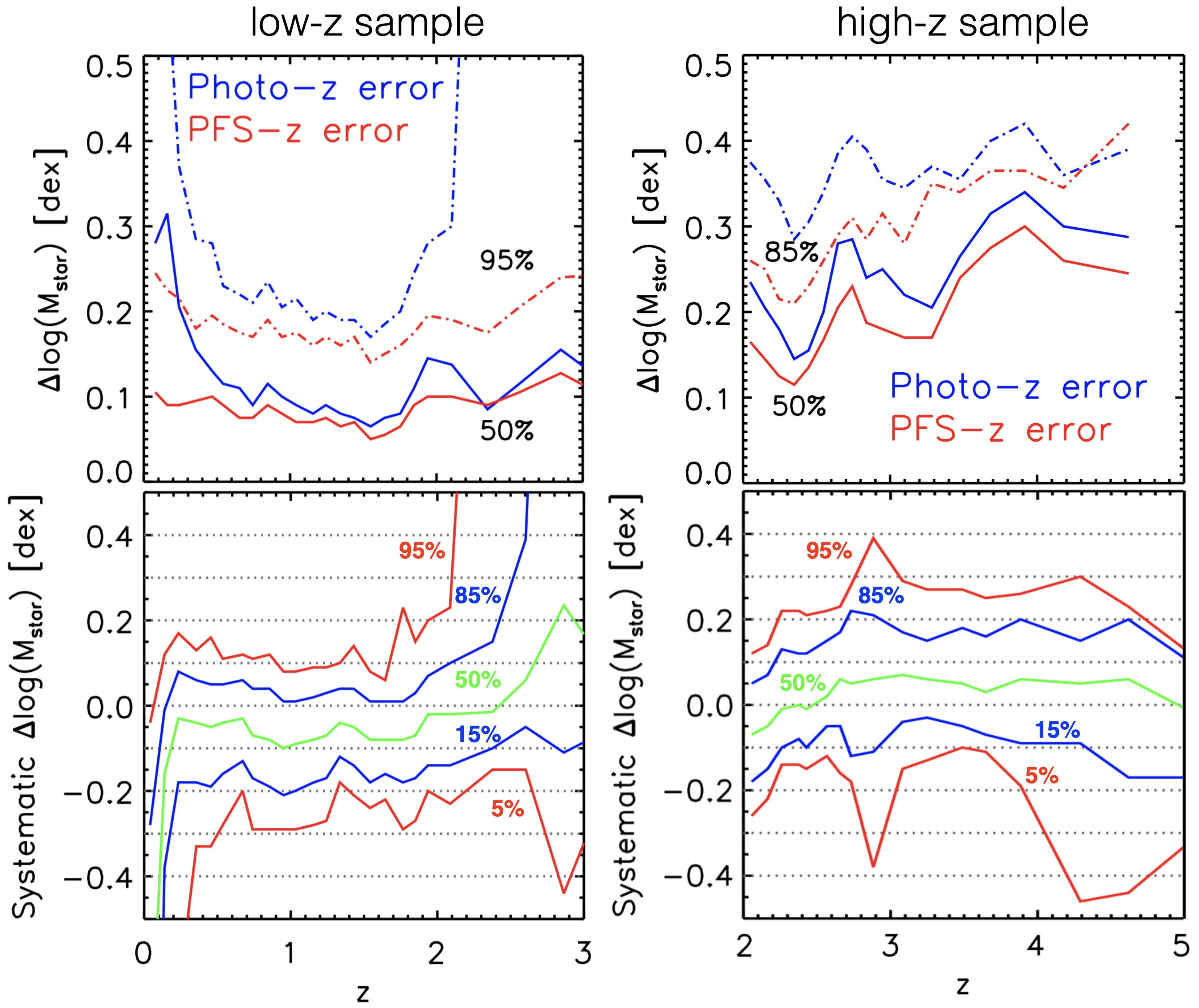}
}
\vskip -0mm
\figcaption[]{
\small
Errors in $M_*$ as a function of redshift for both the $z<2$ continuum-selected (left panels) and the $2<z<4$ drop-out selection (right panels). The top panels show the impact of random errors in the photometric redshifts while the bottom panels show systematic errors. These errors compare the best possible mass with the mass we will achieve in the survey. ''Best'' masses are derived using 50-band COSMOS photometry \citep[from the GALEX UV to the MIPS 24 micron][]{muzzin:2013} and the best available redshift. ``PFS'' masses come from the HSC-Deep photometry ($u, \, grizy, \, JK, \rm{CH12}$, \S \ref{sec:photometry}) and photometric redshifts derived using the HSC-Joint photometry. SED-modeling assumptions were kept the same. The green curve shows the median in the difference, and the blue/red curves shows the 5-95/15-85 percentile curves. Even with the more limited photometry available to us, we will achieve reliable masses for targeting and science.
\label{fig:mtest}}
\end{figure*}

For galaxies with $z{<}2$, fitting the spectral continuum provides significant information about the star formation history. With a combination of PFS spectra and $JK$+\emph{Spitzer} photometry, we can achieve a scatter of $\sim$0.2 dex in stellar mass, comparable to the systematics in stellar population modeling (Figure \ref{fig:mtest}, right) and demonstrably better than typical values based on photometric data alone \citep[e.g, 0.3 dex scatter in stellar mass and 0.4 dex scatter in stellar age;][]{muzzin:2009}.

For galaxies at higher redshifts ($z>2$), the spectra probe only the rest-frame UV, thus the stellar mass estimates are solely based on broad-band photometry, in particular the NIR+\emph{Spitzer} photometry, combined with accurate spectroscopic redshifts from PFS that improve the stellar mass precision over photometric redshifts. We use photometric data with similar depths to our galaxies and redshift precision as set by the redshift simulations to find 95\% of the galaxies have stellar mass precision better than 0.2 dex (Figure~\ref{fig:mtest}). This is substantially better than the photo-$z$ only case, in which a significant tail of objects have unconstrained photo-$z$ at $z>2$. Note that the redshift errors are systematics limited because of velocity shifts between Ly$\alpha$, ISM absorption lines, and the systemic velocity of the galaxy \citep{steidel:2014}.

\subsection{Star Formation Rates}
\label{sec:deliversfr}

We will use our H$\alpha$ measurements out to $z \approx 0.9$ to calibrate star formation rates based on \oii+UV continuum luminosity to apply to the Main sample. We will be sensitive to unreddened star-formation rates of a few tenths of a solar mass per year at $z\sim 0.8$ (based directly on H$\alpha$) and of a few solar masses per year at our median $z=1.5$, based on [O{\small II}]. At $z \sim 1.5$, the unobscured star-formation rate is a few solar masses per year for $L^*$ galaxies, a limit that we can readily reach \citep{Whitaker:2017}. Our effective star formation sensitivity will be a factor of a few lower, given the typical $A_V \sim 1.8$ seen in emission-line galaxies at $z \sim 1.5$  \citep{price:2014,martis:2016}. 

At higher redshifts ($z\gtrsim2$), unobscured star formation will be assessed from the absolute UV magnitude ($M_{UV}$). The similarity of the IRX-$\beta$ relation for local and high-$z$ galaxies up to $z\sim4$ \citep{Fudamoto2020} allows us to account for the obscured component to the SFR in cases for which the UV slope ($\beta$) can be evaluated using both photometry and PFS spectra.

\noindent
\subsection{Deep Spectra: Dynamics and Star Formation Histories} 
\label{sec:deliverdynamics}

\begin{figure*}
\centering
\vbox{ 
\vskip -5mm
\hskip -1mm
\includegraphics[angle=0,width=0.8\textwidth]{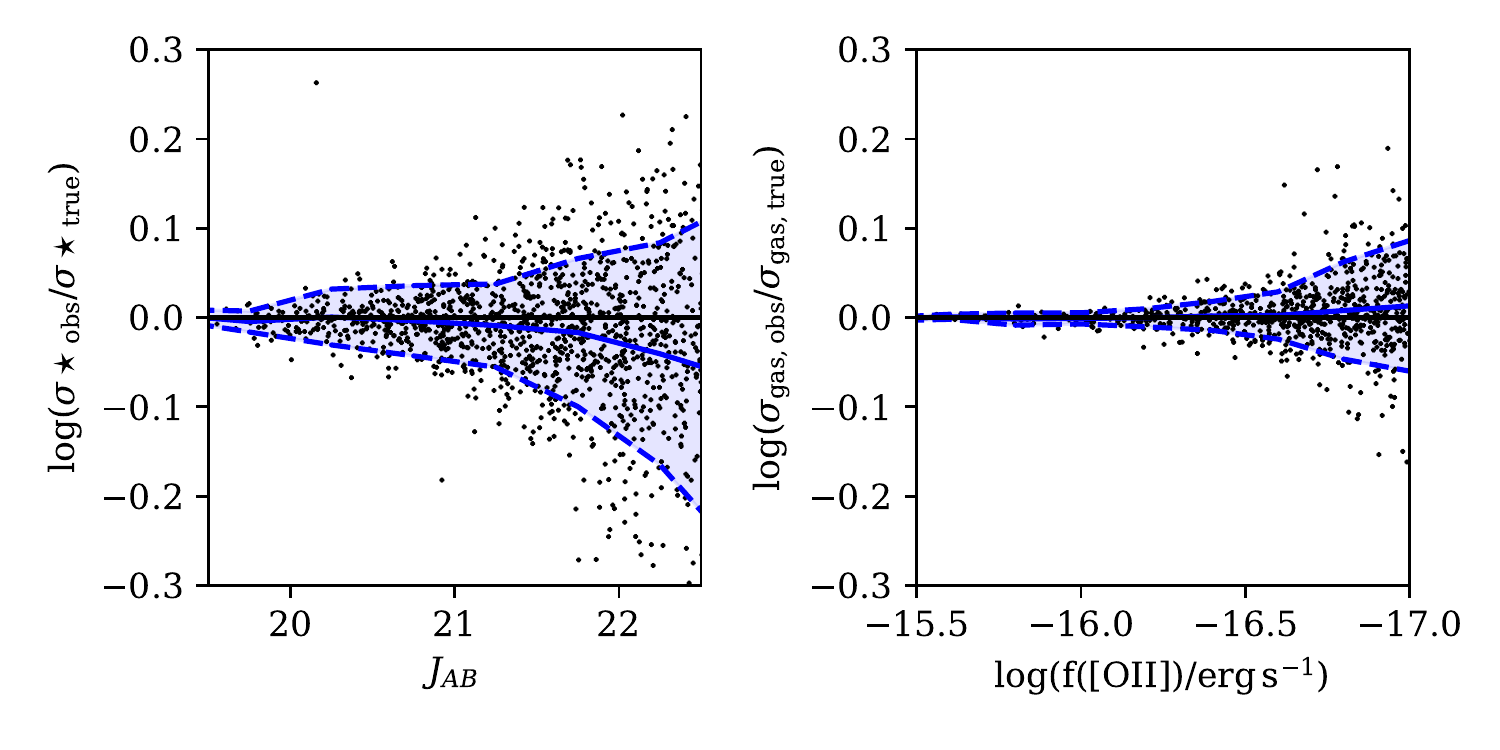}
\vskip -0mm
}
\figcaption[]{
\small
{\it Left}: Recovery of stellar velocity dispersions from 12-hour integrations for the Deep sample. Even at our Deep magnitude limit of $J=22.5$ mag, we will derive meaningful measurements of stellar velocity dispersions for individual galaxies.
{\it Right}: Similar test but for emission line widths. At our detection limit we we still getting reasonable and unbiased line-width measurements of gas emission lines. Recent results \citep{Bezanson:2018} show that the gas velocity dispersion tracks the stellar velocity dispersion without bias, and so we can use gas emission lines as a proxy for stellar velocity dispersion for the main galaxy sample.
\label{fig:deep}}
\end{figure*}

To evaluate our ability to recover higher-order dynamical information from the Deep spectra, we generated 6, 12, and 20 hr integrations with the PFS simulator, and evaluated our ability to recover the key spectral parameters of stellar velocity dispersion and stellar ages. Figure \ref{fig:deep} shows that with 12 hr integrations, we will be able to recover stellar velocity dispersions and stellar ages at our Deep magnitude limit.

For individual galaxies, ${<}15\%$ errors will allow us to derive dynamical masses that are competitive with the stellar population masses. We will achieve this level of precision down to $J=21.5$ mag. Previous studies, e.g., \citealt{Vanderwel:2008}, have explored the size-$\sigma_*$ evolution with only 50 galaxies; with PFS and HSC imaging, we will be able to do this for orders of magnitude more galaxies, and thus explore trends in dynamical mass with star formation history and environment. We will also use these data to calibrate a relation between $\sigma_*$ and gas line width $\sigma_g$, as can be measured at the low redshift end of our sample using LEGA-C \citep{Bezanson:2018}. We will use this relation to estimate $\sigma_*$ for the Main sample, so that we can stack quiescent galaxies in bins of inferred $\sigma_*$ to measure stellar ages and abundance ratios.

We have performed detailed recovery tests of star formation history recovery jointly constrained with simulated 12-hour PFS spectra and the available photometry. The tests include full constraints on a wide range of galaxy properties, including non-parametric input star formation histories with sophisticated models of starbursts and quenching. These tests show that we can recover the average age of a quenched galaxy at z=1.4 with $M_* = 10^{11}$~\msun\ to an accuracy of 4\%, and determine the very low levels of residual star formation in these systems (specific star formation rate $\sim 10^{-11}$\,yr$^{-1}$) to within a factor of two. Similarly, for a bursty star-forming system, we determine the average age of the stars to within 15\% and the recent star formation rate to within 20\%. A key challenge is the age and strength of the burst: the recovery tests show we can infer the time and total mass formed in the recent burst (0.3-1 Gyr in the past) to within a factor of two thanks to the strong continuum constraints from PFS.

\vskip -5mm

\subsection{Gas-phase metallicities and ISM physics}
\label{sec:deliverism}

\begin{figure}
    \centering
\vbox{
\includegraphics[angle=0,width=0.45\textwidth]{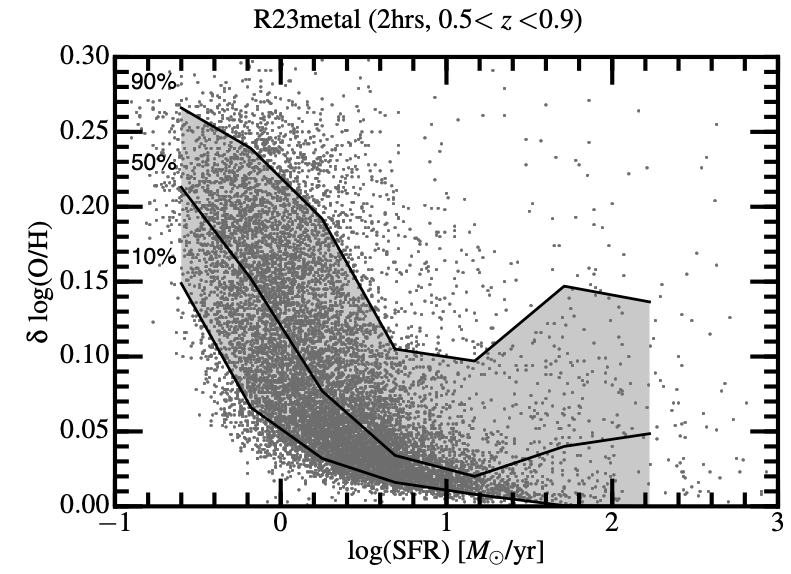}
}
\vskip -0mm
\figcaption[]{
Measurement errors for R23-based gas-phase metallicities (gray dots), estimated from the mock spectra, as a function of SFR.
Solid lines show 90th, 50th, and 10th percentiles of the metallicity measurement errors from top to bottom. 
Pixel noises on the mock spectra are carefully propagated into emission line flux measurement errors and thus metallicity measurement errors.  
Here we assume 2-hour integration with PFS.
\label{fig:metallicity}}
\end{figure}

The wide wavelength coverage of PFS will enable a variety of both restframe-UV and restframe-optical emission line diagnostics of galaxies' enrichment and physical conditions, including gas density and ionization state. Specifically, gas-phase metallicity (typically O/H) can be measured using the ratios of ``strong" rest-optical lines, including \begin{equation}
	R_{23} = \log\left(\frac{[\textrm{O}\,\textsc{ii}]\lambda\lambda3727,3729+[\textrm{O}\,\textsc{iii}]\lambda\lambda4959,5007}{\textrm{H}\beta}\right),
\end{equation}
which will be observable in the spectra of galaxies at $z{<}1.5$. Figure~\ref{fig:metallicity} shows the uncertainty in O/H determined using $R_{23}$ for mock galaxies as a function of SFR, based on the realistic emission-line catalog from the simulation in Section \ref{sec:mock}. Accounting for measurement uncertainties in the emission lines, we expect typical metallicity errors ${<}0.15$~dex (${<}0.25$~dex) for galaxies with $\log$(SFR/($M_\odot$\,yr$^{-1}$))$>0.5$ at $0.5{<}z{<}0.9$ ($0.9{<}z{<}1.5$). 
For galaxies with $z\lesssim0.9$, we will also be able to measure nitrogen-to-oxygen abundance ratios (N/O) using the [N~II]$\lambda\lambda6549,6583$ doublet. Nitrogen is thought to originate via both primary (metallicity-independent) and secondary (metallicity-dependent) production, with N/O increasing with increasing O/H in more chemically mature galaxies \citep[e.g.,][]{vilacostas1993}. As a result, combining measurements of N/O (and C/O, see below) with O/H can provide additional insight into the chemical evolution of individual galaxies in the sample.

In addition to elemental abundances, gas density can be determined using the [O~II]$\lambda\lambda3727,3729$ doublet for galaxies at $z\lesssim2.35$. At $z\lesssim1.5$, [O~II] measurements can be combined with measurements of [O~III]$\lambda4969,5007$ to probe the ionization state of the gas, as well as discriminate between predominately ionization-bounded and density-bounded star-forming regions. Finally, classic restframe-optical diagnostics such as the ``BPT" diagrams \citep[][]{1981PASP...93....5B,veilleux1987} can be used to identify contributions from different sources of ionizing radiation in galaxies, including star formation, AGN (see \S~\ref{sec:deliveragn}), and shocks.

 At higher redshifts, restframe UV emission lines, including Ly$\alpha$ and CIII]$\lambda1909$ \citep[e.g.,][]{Nakajima18,Byler2018}, will allow us to measure other physical properties of high-$z$ galaxies in addition to the metallicity \citep{Faisst2016}, such as the ionization state of the ISM.  These diagnostics will be particularly accessible in spectra having higher S/N in the redshift regime 
 $2{<}z{<}3.5$, given the chosen magnitude limits and integration times. We expect 20 (5)\% of the galaxies at $z=2.5 (5.1)$ to have CIII] emission detected ($>5\sigma$) while 10\% of the galaxies at $z=5.1$ will be detected at $>3\sigma$
\citep{LeFevre17}. Ly$\alpha$ profiles \citep{Verhamme15} encode information on the {\sc Hi} column density and dust distribution that can be assessed with PFS co-added spectra having information on the systemic velocity.

\subsection{Infall and Outflow}
\label{sec:deliveroutflow}

Over the full redshift range spanned by PFS ($0.7 {<} z {<} 7$), we will have both the strong emission (\oii\ at least) and interstellar absorption lines needed to measure the velocity offsets in the interstellar lines. We will search for gas outflow or inflow using the interstellar absorption lines Mg {\small II} and Fe {\small II} in stacks of $\sim 100$ spectra.  

With PFS, we will provide absorption line measurements such as Mg{\sc ii}2796, 2804, {\sc Cii}1335, and Si{\sc ii}. From simulations, we have found that it is possible to recover reliable \ion{Mg}{2} equivalent widths and outflow velocities from stacks of 100 galaxies for the entire redshift range for the main sample galaxies. We insert \ion{Mg}{2} blueshifts into our mock spectra to examine our ability to recover outflow velocities in the face of redshift errors. 
We find that in stacks of $\sim 100$ galaxies, we can reliably recover average outflow velocities to 10\% precision even with realistic redshift errors.

\subsection{Accreting Black Hole Signatures}
\label{sec:deliveragn}

In the first place, we will uncover obscured and low-luminosity active galaxies using emission-line diagnostics on the galaxies in our main continuum-selected sample ($0.7<z<1.7$).  This will allow us to study the factors that may trigger accretion, as well as the impact of the AGN on their hosts \citep[e.g.,][]{kauffmann:2003agn}. We will be able to measure H$\beta$, [O {\small III}], H$\alpha$, and [N {\small II}] for all accreting black holes with $M_{\rm BH}> 10^7$~\msun\ and Eddington ratios ${>}0.05$ at $z{<}0.9$. These lines can also be used to estimate the metallicity of the host galaxy. We will supplement the above sample with luminous sub-populations of active galaxies using multi-wavelength selection. We will measure the black hole mass based on Mg {\small II} and/or H$\beta$, with ${<}$0.2 dex accuracy across $0 {<} z {<} 6$, not including calibration uncertainty. This measurement will be combined with the continuum luminosity to derive the bivariate distribution function of mass and Eddington ratio, the most fundamental quantity to characterize black hole evolution. 

\begin{figure}
\center{\includegraphics[angle=0,width=0.48\textwidth]{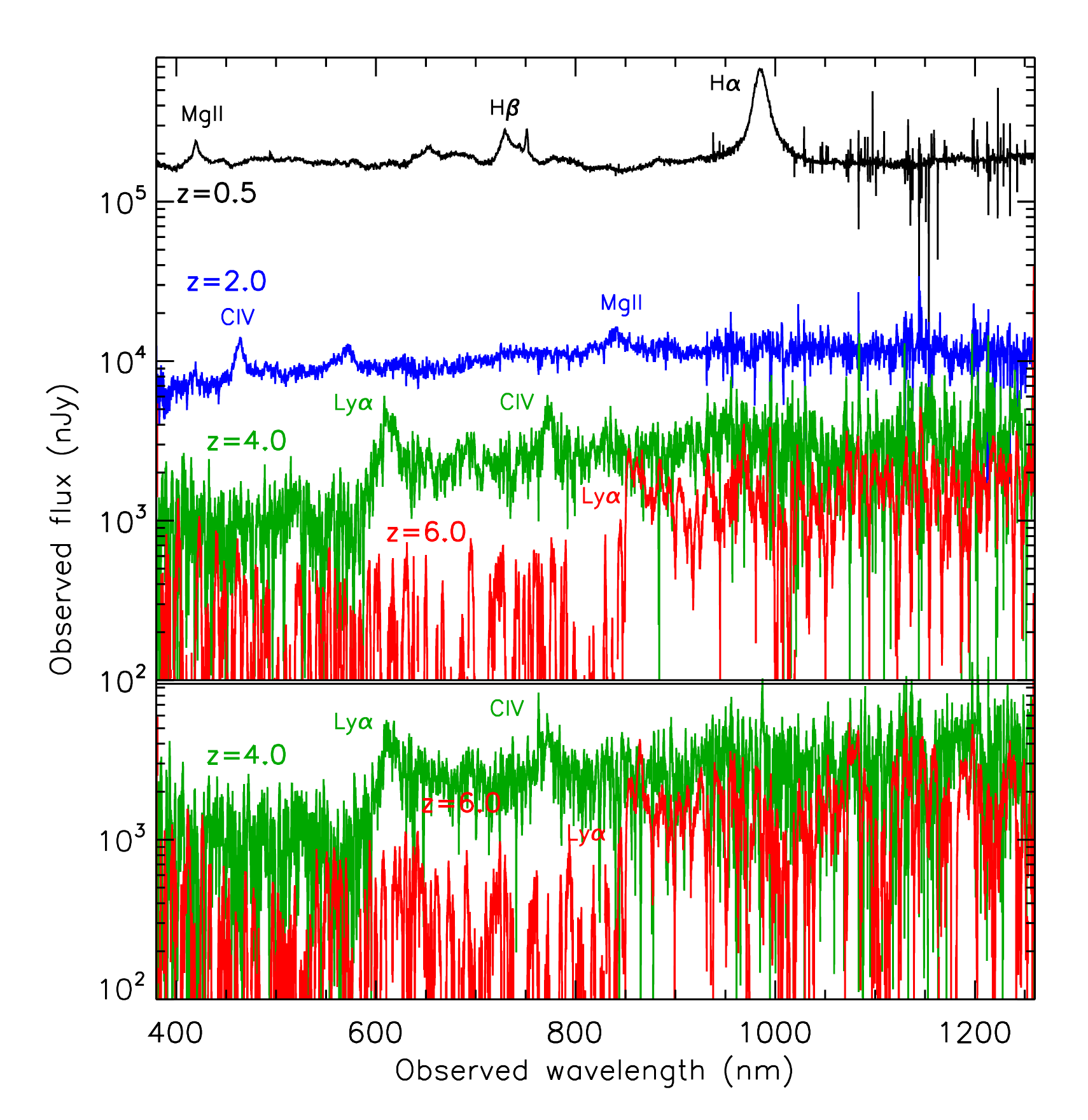}}
\caption{{\it Top}: Simulated PFS spectra of quasars with $M_{1450} = -23$ mag placed at $z = 0.5$ (black), 2.0 (blue), 4.0 (green), and 6.0 (red). The average observing conditions and 1-hr on-source integration are assumed in the simulations.
{\it Bottom}: Simulated 30-min spectra of simulated cosmology quasars placed at z = 4.0 and 6.0. As above, $M_{1450} = -23$ mag.
\label{fig:agn_spec_sim}}
\end{figure}

Detailed modeling of the emission lines will further allow us to measure key physical quantities, such as $M_{\rm BH}$, $\lambda_{\rm Edd}$, 
and gas-phase metallicity.
The wide spectral coverage of PFS enables uniform measurements of $M_{\rm BH}$ with \ion{Mg}{2} $\lambda$2800 at $0.35 {<} z {<} 3.5$
and with \ion{C}{4} $\lambda$1549 at $z > 1.5$, 
and also provides a cross-calibration of the $M_{\rm BH}$ estimates based on the different lines (see Figure \ref{fig:agn_spec_sim}). 
We will exploit the statistical BL sample at $0 {<} z {<} 4$ to measure the 
fundamental, bivariate distribution function of $M_{\rm BH}$ and $\lambda_{\rm Edd}$ with unprecedented accuracy, 
down to $\sim$2 mag lower luminosity 
than previous studies \citep[e.g.,][]{2016A&A...588A..78B}.
The sample at $z > 4$ will be used to measure the faint end of the quasar luminosity function 
\citep[e.g.,][]{2011ApJ...728L..25I, 2012ApJ...756..160I}, as well as the auto-correlation function or cross-correlation function with HSC photometric galaxies, 
which will provide insight into the underlying dark matter halo mass distributions.

The central engines of the X-ray, mid-IR, and sub-mm targets are not necessarily obscured, and thus will partially overlap with the BL sample. 
For obscured objects, PFS will primarily measure emission lines from the narrow line regions and/or the host galaxies, providing accurate redshifts,
source identification, 
and gas kinematics (after stacking if necessary).
This is also the case for radio AGNs \citep[e.g.,][]{2020AJ....160...60Y}.
We will constrain their statistical properties such as number density and clustering amplitudes, which will then be compared to/merged with those of 
the BL sample, in order to reveal the full picture of the cosmological assembly of SMBHs and the host galaxies.
This picture is further complemented by the identification of IMBHs in the local universe; PFS spectra of the IMBH candidates will be examined to find broad 
component in the emission lines, such as H$\alpha$ and H$\beta$, which will then be used to estimate black hole masses.

Finally, for lensed quasar candidates, PFS spectroscopy will provide (i) confirmation of the lensing nature and (ii) accurate redshifts of the 
background sources and/or the foreground lenses.
These pieces of information are essential for most scientific studies exploiting the lensed systems, as partially raised in the previous section.

\subsection{IGM Tomography Recovery}

\begin{figure}
    \centering
    \includegraphics[angle=0,width=0.40\textwidth]{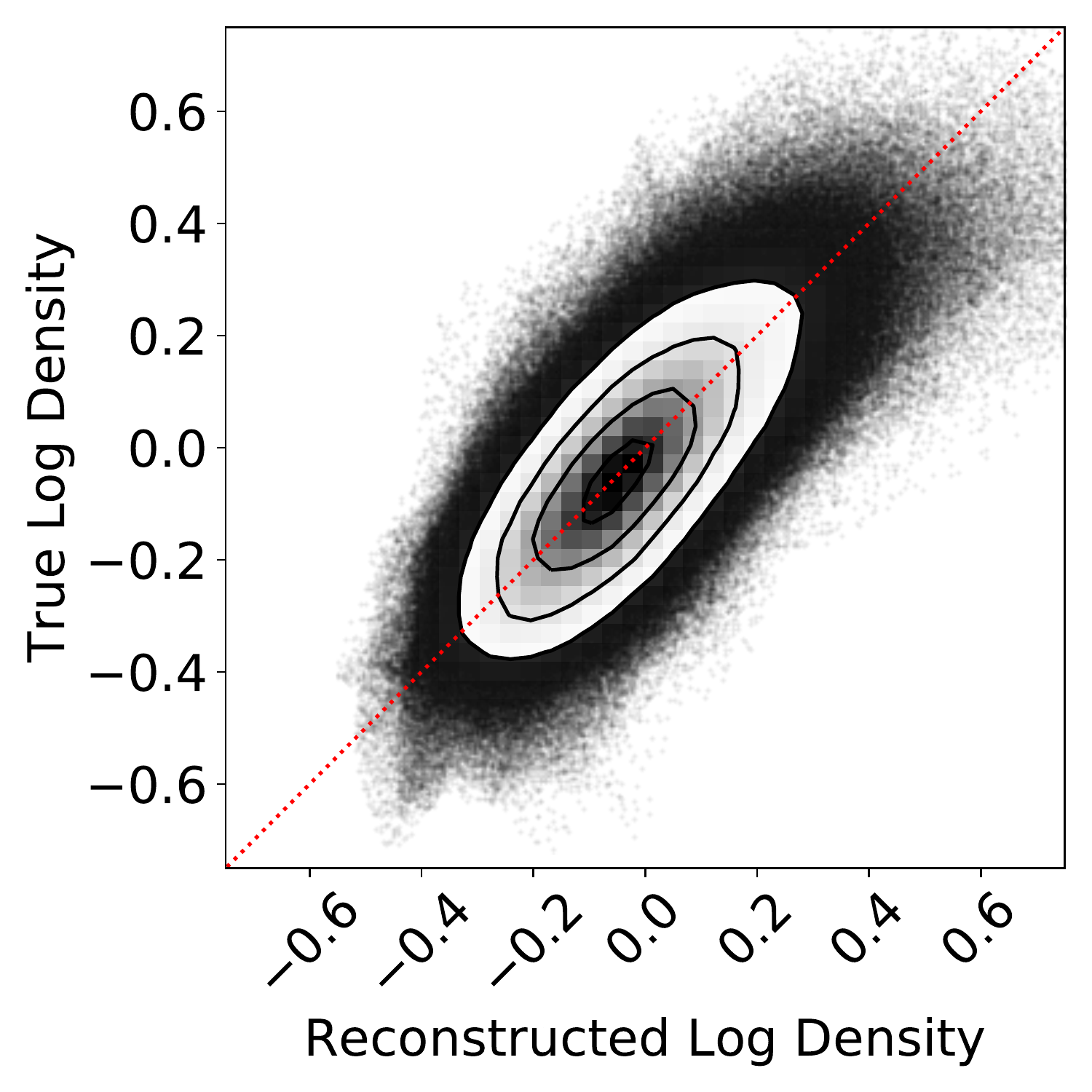}
    \caption{Comparison of reconstructed dark matter density at $z \sim 2.5$ to simulated dark matter density from the Ly$\alpha$ forest absorption together with foreground galaxy positions in a PFS-like mock catalog. The Pearson correlation coefficient between the reconstructed densities and true underlying densities is $r_\mathrm{pearson}=0.72$. \citep{horowitz:2021a}.}
    \label{fig:recon_scatter}
\end{figure}

The survey component at $2.2{<}z{<}3.5$ is primarily designed to reveal the cosmic web at $2.2{<}z{<}2.6$. As such, the main deliverable is the recovery of the underlying 3D matter density field within the targeted
volume. Using the forward modelling TARDIS-II algorithm applied 
jointly to mock Ly$\alpha$ forest absorption data \textit{and} the foreground LBG sample, we find that the PFS GE survey will be able to
recover the smoothed density field (convolved by a $2\,h^{-1}\,\mathrm{Mpc}$ Gaussian kernel) with a Pearson correlation coefficient of $r_\mathrm{pearson}=0.72$ over logarithmic overdensity voxels. The efficacy of this reconstruction can be seen in Figure~\ref{fig:recon_scatter}, where the 3D matter density field in a mock PFS sample is well-recovered over a range of overdensities at $z\sim 2.3$.

\begin{figure*}
    \centering
\vbox{
\includegraphics[angle=0,width=0.9\textwidth]{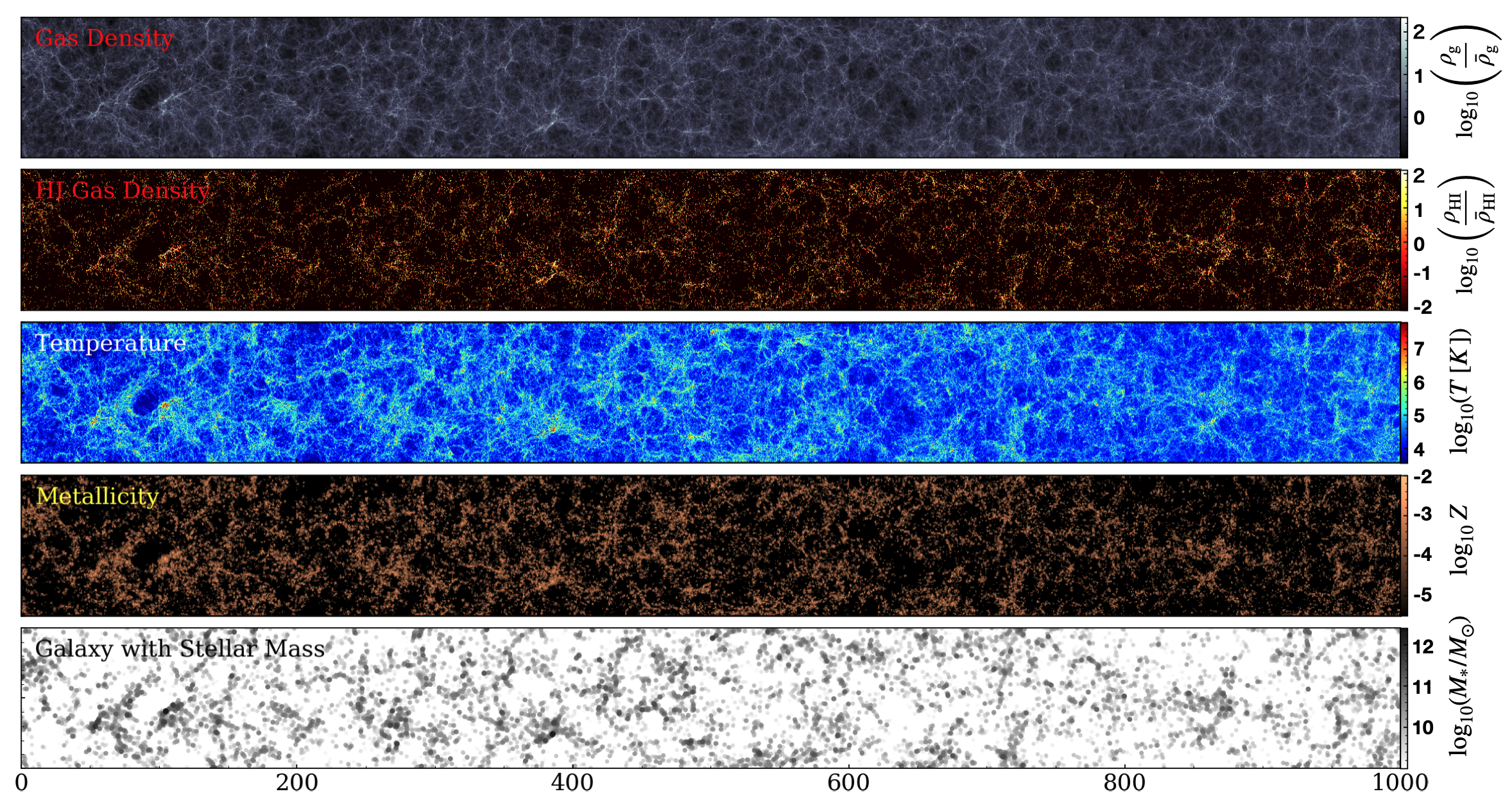}
}
\vskip -0mm
\figcaption[]{
\small
Light-cone output at $z\approx 2-3$, covering $100\,cMpc h^{-1}, {\rm (height)} \times 1\,h^{-1}{\rm cGpc} \times 10\, cMpc h^{-1}$ (depth) from the Osaka group's cosmological simulation.
  Panels from top to bottom show: projected gas overdensity, {\sc Hi} overdensity, temperature, metallicity, and galaxy distribution color-coded by the stellar mass, respectively.
  In the bottom panel, one can see the effect of galaxy bias, where more massive galaxies are more clustered in high-density regions. 
}
\label{fig:Lightcone}
\end{figure*}

\subsubsection{Hydrodynamical simulations of the IGM}

Apart from reconstructions of the cosmic web and matter density field in the IGM tomography sample, one of the major science applications would be the cross-correlation between the IGM absorption and foreground galaxies. 
The interpretation of such cross-correlations, however, would require cosmological hydrodynamical simulations. 
Fortunately, cosmological hydrodynamic simulations have reached a level of sophistication over the past years that the formation of large-scale structure and galaxy formation can be solved together from high-redshift to the present day based on the gravitational instability paradigm within a Cold Dark Matter model. Subgrid models for star formation and feedback allow us to compute the cosmic star formation history and the interaction between galaxies, CGM and IGM.  To facilitate the IGM cross-correlation measurements from PFS GE data, the Osaka group has developed cosmological hydrodynamic simulations of a 100\,cMpc/h box including the models of star formation and supernova feedback 
\citep{Shimizu19,Oku22}, and other box sizes are also being prepared. 
They created light-cone outputs between $z=2-3$  (with evolving density fields and taking peculiar velocities into account) to generate a mock dataset for IGM tomography that was used in designing this survey \citep[see the Appendix of][]{Nagamine21}. 
These simulations can be used to assess the impact of varying feedback models on the IGM tomography and HI--galaxy cross-correlation.

\subsection{Characterizing Groups and Clusters}
\label{sec:deliverlss}

With roughly ten times the area coverage as existing spectroscopic surveys at comparable redshifts, we expect to detect a statistically meaningful number of clusters with $M_{halo}>10^{14}$~\msun, but also enough rich groups to investigate how star formation proceeds in the group environment as a function of their larger-scale environment. The area of the survey is chosen to mitigate cosmic variance and to be large enough to probe rare rich environments, while the sampling rate is high enough to identify individual groups and filaments. Concretely, we expect to find $N_{\rm halo} \approx 2000,500,40$ halos with masses $M_{\rm halo}>10^{13},10^{13.5},10^{14}$~\msun, while we expect to find voids with projected sizes ranging from $10-70$ Mpc (although the exact numbers depend on cosmology). We note that there are no existing void catalogs at $1{<}z{<}1.5$, and we will not be able to detect the large voids at all without $>10$~deg$^2$ of coverage. We will also detect unprecedented numbers of spectroscopically selected protoclusters at $2{<}z{<}6$ \citep[e.g.][]{Tasca2017,Toshikawa2018,Higuchi2019}, allowing crucial systematic tests of cluster growth \citep[][]{chiang:2013} and the onset of environmental effects on galaxy evolution at high redshift \citep[][]{shimakawa2018}. With these new large-scale structure measurements, we expect to uncover new trends. In terms of clustering, we will be able to measure the bias $b_{\rm sub}$ with unprecedented subsamples in $M^*$ and star-formation rate. We will determine whether, at fixed stellar mass, quenched galaxies are preferentially found in denser environments. Going beyond two-point functions, in simulations trends between galaxy specific star formation rate and filament distance (at fixed local galaxy overdensity) are found due to increased gas accretion within the filament.

The powerful new analysis tool of constrained simulations will also be brought to bear on the protoclusters detected by PFS, allowing us to directly model and trace their cosmic evolution in their full large-scale structure context, as they evolve from their observed epochs $>10\,$Gyrs in the past into massive galaxy clusters by our present day (e.g. \citealt{Ata2022}). 
This will allow us to estimate the individual final masses and merger histories of these cosmic superstructures with much better accuracy than feasible with generic cosmological simulations.

\subsubsection{Groups and protocluster candidates from the photometric redshift data}
\label{sec:group_photo}

To demonstrate the power of the CLAUDS+HSC photometry to identify proto-cluster candidates, group and protocluster candidates were recently obtained by \citet{Li2022} from 5.6 million galaxies with $i$-band magnitude $i {<} 26$ within the redshift range ($0{<}z{<}6$)\footnote{Data are available at this link: \url{https://gax.sjtu.edu.cn/data/PFS.html}}. Here the groups were identified using the extended halo-based group finder developed by \citet{Yang2021}, which is able to deal with galaxies via spectroscopic and photometric redshifts simultaneously.  A total of 2.2 million groups were identified, within which 400k groups have at least three member galaxies. By checking the galaxy number distributions within a $5-7\hMpc$ projected separation and a redshift difference $\Delta z \le 0.1$ around those richest groups at redshift $z>2$, we have obtained a list of 761, 343 and 43 protocluster candidates in the redshift bins $2\leq z{<}3$, $3\leq z{<}4$ and $z \geq 4$, respectively.  These groups and protocluster candidates may suffer from the photometric redshift errors, but PFS will be very effective at confirming them and identifying the brightest cluster galaxy as well.

\subsection{Mapping the Cosmic Web}
\label{sec:cosmicwebmap}

In conjunction with HSC, PFS will be able to not only constrain the intrinsic alignment of galaxy shapes with the underlying cosmic web but explore this relationship as a function of redshift (see Figure \ref{fig:LSS} for examples of our ability to reconstruct the cosmic web). To compare observed galaxy shapes with the underlying cosmic web, we first reconstruct the three-dimensional matter field from observed galaxy positions and/or \lya\ absorption using either a dynamical forward modelling approach \citep{2019TARDIS,2021TARDISII} or a Wiener filter \citep{pichon01a}. 
In addition to these, we will also use the galaxy groups to construct the matter field \citep{Wang2012}.
From these three-dimensional reconstructions we calculate the Hessian of the inferred potential of the field, known as the deformation tensor. The signature of the eigenvalues of this tensor can be used to classify the cosmic structures (i.e., void, sheet, filament, node) and the eigenvectors can be used to define their spatial orientation \citep{lee_white16,krolewski17a}. The principle spatial orientation of the underlying cosmic web can then be correlated to the inferred shape and/or spin of the observed galaxies to constrain the intrinsic alignment \citep{Zhang2013, Zhang2015}. This analysis can be performed both in the continuum-selected samples using galaxy-only reconstruction and in the IGM tomography sample using joint galaxy-\lya Forest reconstruction. 

\section{Summary}

The PFS Galaxy Evolution Survey will change our view of galaxy evolution by studying typical galaxies over most of cosmic time, from the epoch of reionization until $z \sim 0.7$. The wide wavelength coverage provided by the spectrograph will allow for detailed studies of star formation histories, abundances and abundance ratios, the prevalence of outflows, and so much more, while the high multiplexing capability means that we will have unprecedented sampling with which to reconstruct the cosmic web. We have built a survey to take advantage of these strengths, and to complement other upcoming work in this decade.

\clearpage

\newpage

\bibliography{master.bib}

\end{document}